\documentclass[twocolumn,amsmath,amssymb,floatfix,superscriptaddress,prb,longbibliography]{revtex4-2}
\usepackage{color,soul}
\usepackage {graphicx}
\usepackage {epstopdf}
\usepackage{chemfig}
\usepackage{lineno,hyperref}
\usepackage{bm}
\usepackage{amsmath}
\usepackage{gensymb}
\usepackage{epstopdf}
\usepackage{booktabs}
\usepackage{longtable}
\usepackage{tabularx}
\usepackage{multirow}
\usepackage{setspace}
\usepackage{mathrsfs}
\usepackage{makecell}

\begin{document}
\title{From a High-T$\rm_C$ Spin-Gapless Semiconductor to a Spin-Glass: The Role of Disorder in NiMnCrAl}

\author{Arka Patra}
\affiliation{Condensed Matter and Surface Physics Division, Saha Institute of Nuclear Physics,
	A CI of Homi Bhabha National Institute, 1/AF Bidhannagar, Kolkata 700064, India}
\author{Tufan Roy}
\affiliation{Centre for Science and Innovation in Spintronics, Tohoku University, Sendai  980-8577, Japan }
\affiliation{Department of Physics, School of Advanced Sciences (SAS),Vellore Institute of Technology, Vellore 632014, Tamil Nadu, India}
\author{Nilofar Kurawle}
\affiliation{UGC-DAE Consortium for Scientific Research, Mumbai Centre, BARC, Trombay, Mumbai 400085, India}
\author{Tamaghna Maitra}
\affiliation{Condensed Matter and Surface Physics Division, Saha Institute of Nuclear Physics,
	A CI of Homi Bhabha National Institute, 1/AF Bidhannagar, Kolkata 700064, India}
\author{Souvik Banerjee}
\affiliation{Chemistry and Physics of Materials Unit, School of Advanced Materials, JNCASR, Bengaluru 560064, India}
\author{Mantu Modak}
\affiliation{Condensed Matter and Surface Physics Division, Saha Institute of Nuclear Physics,
	A CI of Homi Bhabha National Institute, 1/AF Bidhannagar, Kolkata 700064, India}
\author{Shuvankar Gupta}
\affiliation{Condensed Matter and Surface Physics Division, Saha Institute of Nuclear Physics,
	A CI of Homi Bhabha National Institute, 1/AF Bidhannagar, Kolkata 700064, India}
\author{Sudip Chakraborty}
\affiliation{Condensed Matter and Surface Physics Division, Saha Institute of Nuclear Physics,
	A CI of Homi Bhabha National Institute, 1/AF Bidhannagar, Kolkata 700064, India}
\author{Soumya Bhowmik}
\affiliation{Condensed Matter and Surface Physics Division, Saha Institute of Nuclear Physics,
	A CI of Homi Bhabha National Institute, 1/AF Bidhannagar, Kolkata 700064, India}
\author{Papri Dasgupta}
\affiliation{Physics Department, Jadavpur University, Kolkata 700032}
\author{Rajashri Urkude}
\affiliation{Beamline Development \& Application Section, Bhabha Atomic Research Centre, Trombay, Mumbai 400085, India}
\author{Biplab Ghosh}
\affiliation{Beamline Development \& Application Section, Bhabha Atomic Research Centre, Trombay, Mumbai 400085, India}
\author{D. Bhattacharyya}
\affiliation{Atomic \& Molecular Physics Division, Bhabha Atomic Research Centre, Mumbai 400 094, India}
\author{A. Sundaresan }
\affiliation{Chemistry and Physics of Materials Unit, School of Advanced Materials, JNCASR, Bengaluru 560064, India}
\author{Sudhindra Rayaprol}
\affiliation{UGC-DAE Consortium for Scientific Research, Mumbai Centre, BARC, Trombay, Mumbai 400085, India}
\author{Masafumi Shirai}
\affiliation{Center for Science and Innovation in Spintronics, Tohoku University, Sendai  980-8577, Japan}

\author{Chandan Mazumdar}
\email{chandan.mazumdar@saha.ac.in}
\affiliation{Condensed Matter and Surface Physics Division, Saha Institute of Nuclear Physics,
	A CI of Homi Bhabha National Institute, 1/AF Bidhannagar, Kolkata 700064, India}
	
\begin{abstract}
We combine first-principles calculations with structural, magnetic, and transport measurements to investigate the influence of chemical disorder on the quaternary Heusler alloy NiMnCrAl. For the ordered Y-type structure, density-functional-theory calculations predict a ferrimagnetic spin-gapless-semiconducting state with a total magnetic moment of $1.99\,\mu_{\mathrm{B}}/\mathrm{f.u.}$ and a mean-field ordering temperature of approximately $1300~\mathrm{K}$. Polycrystalline NiMnCrAl forms in a single average cubic phase; however, combined x-ray diffraction, neutron diffraction, and extended x-ray absorption fine-structure measurements reveal substantial antisite disorder, characterized by predominantly B2-like disorder with additional cross-sublattice mixing. No long-range ferrimagnetic transition is observed up to $700~\mathrm{K}$. Instead, the compound exhibits a broad magnetic anomaly near $37~\mathrm{K}$, frequency-dependent ac susceptibility, slow stretched-exponential relaxation, and field-cooled and zero-field-cooled memory effects, establishing a glassy magnetic state. The electrical resistivity is metallic and the Seebeck coefficient is negative, consistent with the loss of the ideal spin-gapless electronic structure. Exchange-interaction calculations show that antisite disorder introduces short Mn--Mn separations and competing antiferromagnetic Mn--Mn and Mn--Cr interactions, providing a microscopic basis for magnetic frustration. These results demonstrate the strong sensitivity to the predicted spin-gapless and ferrimagnetic states of NiMnCrAl to chemical disorder.
\\
\\keywords: Magnetism, Heusler, Spin-Glass.
\end{abstract}

\maketitle
\section{Introduction}

Spintronics is an emerging area of research utilizing both the charge as well as spin of the electron for data transport and signal processing, in contrast to traditional electronics which is restriced to charge transport only \cite{Fert}. Practical applications have already started to manifest in many different technologies such as magnetic random-access memory (MRAM) \cite{S.D.Sharma}, advanced read-head sensors for hard drives \cite{Atsufumi Hirohata}, prototype spin-based logic devices \cite{SP Das} and many more. Compounds such as PbPdO$_2$ \cite{wang}, Co$_3$Sn$_2$S$_2$ \cite{SGSGupta2} etc. often offer clear benefits over standard electronic systems, \textit{viz.}, in the field of non-volatile memory, faster switching, and reduced power demand \cite{Atsufumi Hirohata}. Yet, the function of many such materials are still marred with multiple difficulties such as poor spin-injection efficiency, rapid spin-relaxation, and the challenge of combining spintronic components with standard semiconductor circuits. To address these issues, researchers have turned their attention toward half-metallic ferromagnets, which are a unique class of materials where one spin-band conducts electrons like a metal while the other behaves as a semiconductor offering 100~\% electronic spin polarization \cite{Groot}. Some examples include CrO$_2$ \cite{I. V. Solovyev}, $\mathrm{La_{1-x}Sr_{x}MnO_{3}}$ (LSMO)\cite{G. Banach} and a few Heusler compounds such as NiMnSb \cite{Groot}. However, in spite of exhibiting half-metallic characters, such compounds nonetheless offer only rather restricted electronic mobility along with band tuneability necessary for the practical device applications.

The recent discovery of spin-gapless semiconductors (SGS) \cite{wang}, where one spin-channel shows a zero band gap while the opposite channel has a finite gap, provided a great opportunity to address the above-mentioned technological limitations as the overall carrier mobility in SGS compounds turned out to be  considerably higher \cite{Z. Yue} in comparison to those of traditional semiconductors. Additionally, their unusual electronic structure can be easily tuned with doping, application of external electric/magnetic fields, or even  strain. Only a small number of such materials have been experimentally realized so far, which are mostly 3\textit{d} transition-metal based Heusler alloys having valence electron counts 26 or 28,\cite{C. Felser} \textit{viz.}, CoFeMnSi \cite{L. Bainsla}, CoFeMnSn \cite{Gupta2}, and a few others.

Full-Heusler Alloys are a class of intermetallic compound, ideally forming in ordered $L2_1$-type crystal structure having a general formula X$_2$YZ \cite{Gupta 3}, where X and Y are \textit{d}-block elements and Z belongs from the \textit{p}-block of the periodic table. Gradually, the Heusler alloy family have been further expanded into several sub-categories \textit{viz.}, Half-Heusler (XYZ), Inverse-Heusler (Y$_2$XZ) and Anti-Heusler (Z$_2$XY) alloys \cite{Soumya}. Many members of this family of materials are known to host a wide-ranging spectrum of properties; from thermoelectricity \cite{thermoelectric} and superconductivity \cite{superconductivity} to shape-memory behavior \cite{shapememory} and magnetic skyrmions \cite{skyrmion} and hence found their usefulness in various branches of science, including physics, chemistry, material science etc., often blurring their conventional boundaries. On top of it, the unique hybridization in the \textit{d}- and \textit{p}- orbitals in these materials may poise to support many interesting features, such as  Half-Metallic-Ferromagnetism (HMF) \cite{Groot}, Bipolar Magnetic Semiconducting (BMS) \cite{bms} and Spin-Gapless Semiconducting (SGS) \cite{L. Bainsla} characteristics. It was further found that the scope of band-tunability could be expanded to a much greater extent when one of the X-atoms in X$_2$YZ type traditional full-Heusler alloy are replaced by another \textit{d}-block element, X$^\prime$. Such compounds having general formula XX$^\prime$YZ are another, relatively new, sub-class of Heusler alloys known as Quaternary Heusler alloys.

Among the half-metallic full-Heusler alloys, Co$_2$FeSi are known to exhibit highest Curie temperature (T$\rm_C \sim$1100~K) \cite{Co2FeSi1100K,Co2FeSi-1200K}. Such a large value of T$\rm_C$ is highly beneficial for many practical applications that require operational utility over a wide temperature region. When Co is replaced by Mn, the resultant compound Mn$_2$FeSi turned out to be an inverse Heusler antiferromagnet \cite{Mn2FeSiAntiferro}. Very interestingly, the intermediate quaternary composition, CoMnFeSi, though brings down the Curie temperature to 763 K, introduces the coveted spin-gapless semiconduction feature \cite{L. Bainsla}. As the known spin-gapless Heusler alloys are very few in number and most of their respective Curie temperatures do not exceed much beyond that of CoMnFeSi, the realization of new spin-gapless materials with even larger Curie temperatures provides a great opportunity to the research community.

Recently, a theoretical study has identified NiMnCrAl as a promising candidate for a spin-gapless semiconductor, exhibiting ferrimagnetic spin arrangements of Mn, Cr and Ni \cite{NMCANepalT,NMCAGhellabT}. A preliminary experimental investigation also suggested the ferrimagnetic spin arrangement with high transition temperature ($>$400~K), in addition to a peak in magnetic susceptibility around $\sim$37~K. However, the spin-gapless feature of the system was not investigated. In the present work, we have investigated the structural and magnetic properties of polycrystalline NiMnCrAl through x-ray diffraction (XRD), EXAFS, temperature-dependent neutron diffraction, ac- and dc-magnetic susceptibility, isothermal magnetisation, non-equilibrium magnetic dynamics, resistivity, thermal conductivity and Seebeck effect. Appropriate theoretical calculations are also utilized in understanding the experimental results.

\section{Method of Investigation}

\subsection{Theory}

First-principles calculations were performed using full-potential spin-polarized-relativistic Korringa-Kohn-Rostoker (FP-SPRKKR) method, as implemented in the SPR-KKR program package \cite{sprkkr}. We used coherent potential approximation (CPA) to model the substitutional disorder in NiMnCrAl alloy \cite{cpa1,cpa2}. For the exchange correlation functional the generalized-gradient-approximation was used as parameterized by Perdew, Burke, and Ernzerhof. \cite{pbe} The angular momentum expansion of the basis set was limited to $l_{max}=3$. For the determination of Fermi energy, we used Lloyd’s formula \cite{lloyd1, lloyd2}. A $k$-mesh of 33$\times$33$\times$33 was used for the Brillouin zone integration. Furthermore, we evaluated the Curie temperature using mean-field-approximation, following the formalism adopted by Liechtenstein \cite{liechtenstein}.

\subsection{Experimment}

Polycrystalline NiMnCrAl was synthesized by arc melting technique. Constituent elements (M/s Alfa Aesar, $\ge$ 99.99 \%) were taken in 1:1:1:1 ratio in a water-cooled copper hearth with a 2\% additional Mn to account for its high volatility during the melting process. The ingot was melted several times under argon atmosphere to ensure uniformity and the total weight loss during the melting process estimated to be approximately 1.1\%. A part of the resulting ingot was annealed under vacuum at 800$^{\circ}$ C for seven days to look for any possible improvement in structural arrangement. For magnetic and transport measurements, the ingot was sliced into appropriate long, rectangular shapes. X-ray diffraction (XRD) measurement was carried out on powdered sample at room temperature in a TTRAX-III diffractometer (M/s. Rigaku, Japan) using Cu \textit{K$\alpha$} radiation. Both DC and AC magnetic characterizations, along with thermal and electrical transport measurements, were performed using a Physical Property Measurement System (PPMS, M/s Quantum Design Inc., USA) in the temperature range 2 $<$ T $<$ 400~K, magnetic field (H) range -70 kOe $<$ H $<$ +70 kOe and frequency (f) range 10 $<$ f $<$ 10000 Hz. As the basic structural and magnetic properties found essentially unaltered for both the as-cast as well as annealed samples, all the detailed measurements were carried out on annealed samples only. The magnetic susceptibility measurement in the temperature (T) range 300~K $<$ T $<$ 700~K was carried out using the high temperature Vibrating Sample Magnetometer (VSM) setup (M/s. Microsence, USA). Neutron diffraction measurement was performed 300~K in Dhruva PD-2 Diffractometer at Bhabha Atomic Research Centre (BARC), Mumbai, India. The X-ray Absorption Spectroscopy (XAS) measurements, comprising X-ray Absorption Near Edge Structure (XANES) and Extended X-ray Absorption Fine Structure (EXAFS), were carried out at the Mn and Cr K-edges. Data were collected at the Scanning EXAFS Beamline (BL-9) of the Indus-2 Synchrotron Source (2.5 GeV), Raja Ramanna Centre for Advanced Technology (RRCAT), Indore, India. The beamline operates in the 4-30 keV range. Its principal optics include an Rh/Pt coated collimating meridional cylindrical mirror and a double crystal monochromator (DCM) with Si (111) crystals. The collimated polychromatic beam was monochromatized by the Si (111) DCM, giving a resolution ($\Delta E/E$) of $10^{-4}$ at 10 keV. The second DCM crystal was detuned to suppress higher harmonics. Further beamline details are available elsewhere \cite{EXAFS-1}.

\begin{table}
	\centering
	\caption{ Types of possible non-degenerate atomic distributions in the compound NiMnCrAl}
	\begin{tabular}{ccccc}
		\hline
		\hline
		Structure Types  & \hspace{0.5cm} $ 4a $  & \hspace{0.5cm} $ 4b $  & \hspace{0.5cm} $ 4c $ & \hspace{0.5cm} $ 4d $ \\
		
		\hline
		Type - I    & \hspace{0.5cm} Al & \hspace{0.5cm} Mn & \hspace{0.5cm} Ni & \hspace{0.5cm} Cr \\
		Type - II   & \hspace{0.5cm} Al & \hspace{0.5cm} Cr & \hspace{0.5cm} Mn & \hspace{0.5cm} Ni \\
		Type - III  & \hspace{0.5cm} Al & \hspace{0.5cm} Ni & \hspace{0.5cm} Mn & \hspace{0.5cm} Cr \\
		\hline
		\hline
	\end{tabular}
	\label{tab:Structure_types}
\end{table}

\section{Results and Discussions}

\subsection{Ideal structural and magnetic properties}

\begin{table*}
	\centering
	\caption{Relative energies and magnetic moments for different structures-types in NiMnCrAl.}
	\begin{tabular}{ccccccc}
		\hline
		\hline
		Structure & \hspace{0.5cm} Energy (meV/atom) & \hspace{0.5cm} Ni ($\mu_{\mathrm{B}}$) & \hspace{0.5cm}	Mn ($\mu_{\mathrm{B}}$) & \hspace{0.5cm} Cr ($\mu_{\mathrm{B}}$) & \hspace{0.5cm} Al ($\mu_{\mathrm{B}}$) & \hspace{0.5cm} Total ($\mu_{\mathrm{B}}$) \\
		\hline
		Type-I & \hspace{0.5cm} 0   & \hspace{0.5cm} 0.59 & \hspace{0.5cm} 2.95  & \hspace{0.5cm} -1.50 & \hspace{0.5cm} -0.06 & \hspace{0.5cm} 1.99 \\
		Type-II & \hspace{0.5cm} 16  & \hspace{0.5cm} 0.64 & \hspace{0.5cm} -1.18 & \hspace{0.5cm}  2.36 & \hspace{0.5cm} -0.01 & \hspace{0.5cm} 1.81 \\
		Type-III & \hspace{0.5cm} 238 & \hspace{0.5cm} 0.19 & \hspace{0.5cm}  3.07 & \hspace{0.5cm} -1.95 & \hspace{0.5cm} -0.04 & \hspace{0.5cm} 1.27 \\
		\hline
		\hline
	\end{tabular}
	\label{tab:magnetic_moments}
\end{table*}

Quaternary Heusler alloys (XX$^\prime$YZ) usually crystallize in the cubic space group $F\bar{4}3m$ (Space Group No. 216), which can be visualized as four interpenetrating $fcc$-type sublattices, where the four different atoms occupy four different Wyckoff positions, \textit{viz.}, $4d$ (0.75,0.75,0.75) \& $4c$ (0.25,0.25,0.25) sites having tetrahedral symmetry and $4b$ (0.5,0.5,0.5) \& $4a$ (0,0,0) sites having octahedral symmetry. Such an atomic arrangement generally known as Y-type ordered structure  \cite{T Graph}. Depending on the respective choices of X, X$^\prime$, Y and Z atoms to get arranged among these four crystal sites, the structure can be further classified into three clearly distinguishable variants, namely Type-I, Type-II, and Type-III. All the three possible atomic arrangements for NiMnCrAl are presented in Table \ref{tab:Structure_types}.

\begin{figure}
	\includegraphics[width=8.5cm]{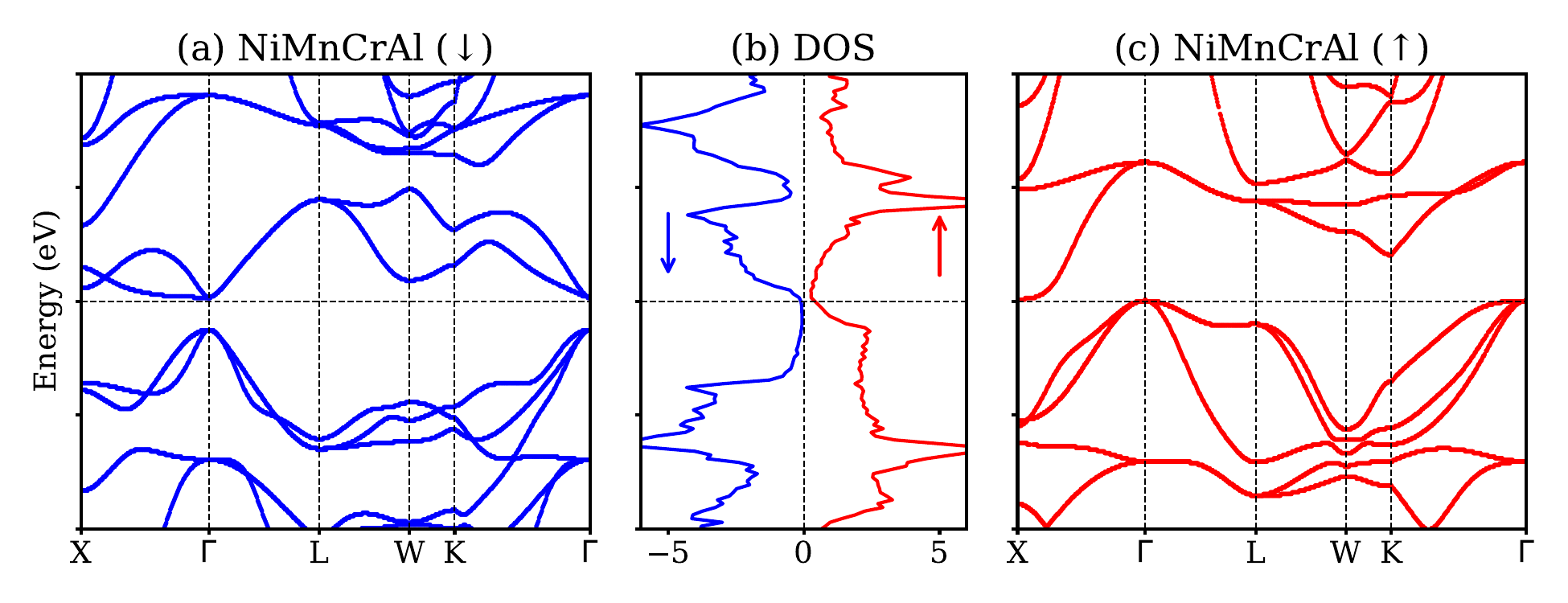}
	\caption {Calculated band dispersion curves and density of states (DOS) for NiMnCrAl. The spin-down DOS is shown in the left panel(a),in the middle panel (b) the total (spin-up + spin-down) DOS is presented, and the right panel (c) shows the spin-up DOS.}
	\label{fig:DOST1}
\end{figure}

To check the relative stability of these crystal structures and the distribution of magnetic moments among different constituent atoms, we have carried out \textit{ab-initio} calculations and the results are presented in Table~\ref{tab:magnetic_moments}. Our calculation reveals that Type-I structure is the energetically most favorable structure in the ordered configuration and have an antiparallel spin alignment between Mn and Cr moment. In the Type-I configuration the magnetic moment is largely localized at the Mn (2.95~$\mu_\mathrm{B}$) and Cr (-1.5~$\mu_\mathrm{B}$), whereas Ni (0.59 $\mu_\mathrm{B}$) carries a tiny moment, resulting a total magnetic moment of $\sim$2 $\mu_\mathrm{B}$ per fourmula unit. This result is consistent with the well established Slater-Pauling (S-P) rule \cite{galanakis-prb-2002} which dictates the magnetic moment of a Heusler system to be expressed as N$_V$-24, where N$_V$ is the total valence electron count (VEC) of the said Heusler alloy. In case of NiMnCrAl, N$_V$ can be estimated as $(3d^{8} + 4s^{2})_{\mathrm{Ni}}$ + $(3d^{5} + 4s^{2})_{\mathrm{Mn}}$ + $(3d^{5} + 4s^{1})_{\mathrm{Cr}}$ + $(3s^{2} + 3p^{1})_{\mathrm{Al}}$ = 26 and hence the theoretical estimation of 2 $\mu_\mathrm{B}$ follows the S-P rule quite well. Our calculation also reveals that this configuration is capable of offering a very high ordering temperature of $\sim$1300~K, exceeding even all the reported Co-based Heusler alloys that are known for their high Curie temperatures and hence would have immense technological importance.The distinctness of the compound is further manifested in its spin polarized band structure (Fig.~\ref{fig:DOST1}) reflecting the rare occurrence of spin gapless semiconducting phenomena where an insulating down-spin channel with a direct band gap of $\sim$0.3 eV is observed at the zone centre  ($\Gamma$ point). On the other hand, in the up-spin channel, top of the valence band and bottom of the conduction band touches the Fermi level at the $\Gamma$ and $X$ points, respectively, leading to a band gap of almost zero width,  conforming the earlier reported theoretical investigation \cite{NMCAGhellabT,NMCANepalT,CrNiMnAl-Published1}. We have extended the calculation for Type-II and Type-III structure as well, where the total magnetic moment estimated to be 1.81 $\mu_\mathrm{B}$ and 1.27 $\mu_\mathrm{B}$, respectively. However, as both Type-II and Type-III structures are energetically unfavorable compared to Type-I, we therefore restrict our subsequent discussion to the Type-I configuration only.

\subsection{X-Ray Diffraction measurement}
\label{sec:XRD}

\begin{figure*}
	\centering
	\includegraphics[width=8cm]{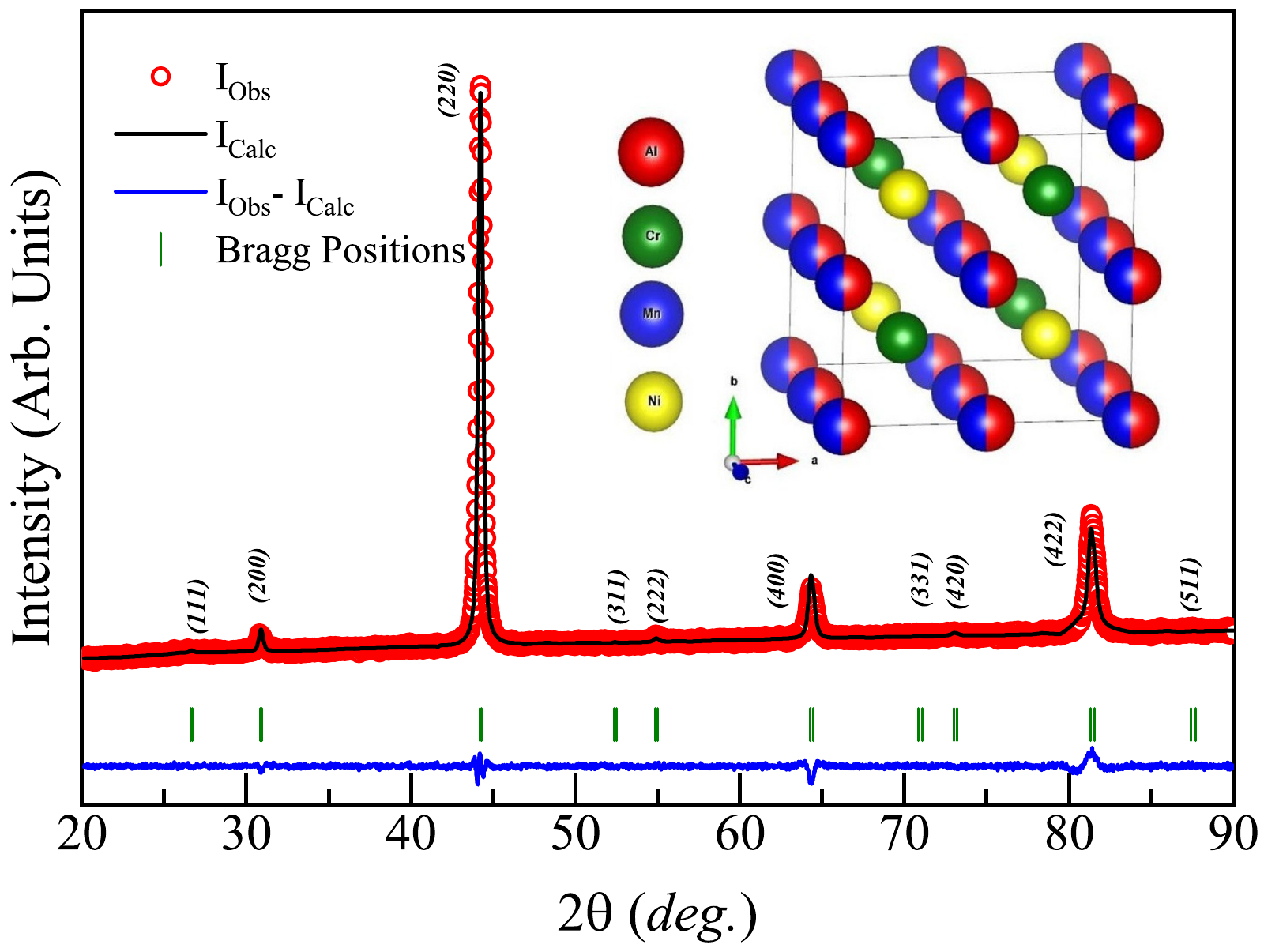}  \includegraphics[width=8cm]{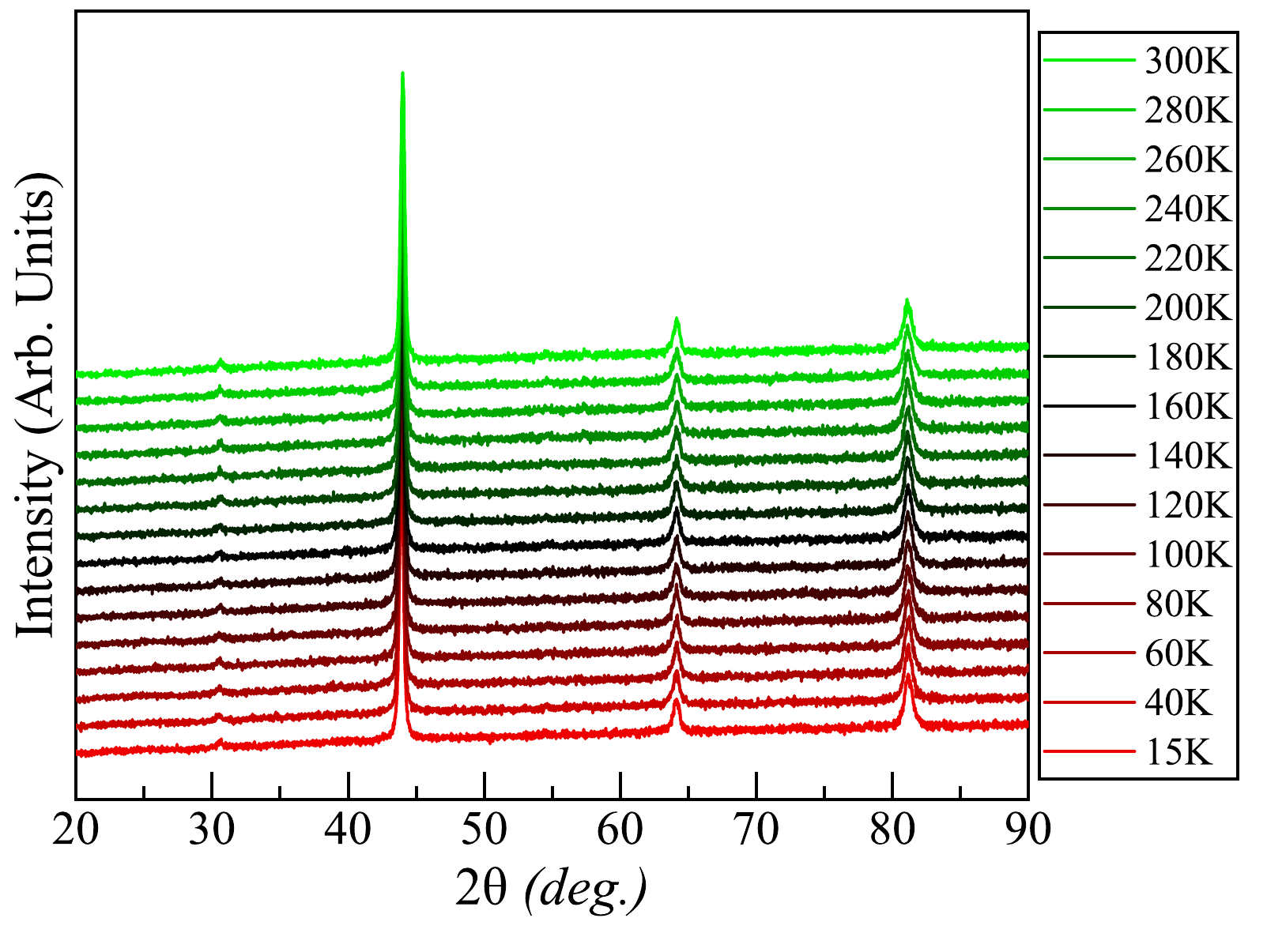}
	\caption{(Left) Rietveld Refinement of the powdered X-Ray diffraction (XRD) pattern using a disordered structure. The equal mixing of Al and Mn in 4\textit{a} and 4\textit{b} site matches the refinement with full stoichiometry. (Inset) The structural model obtained from XRD analysis. This model has been further refined during neutron diffraction analysis, presented in Section \ref{sec:Neutron} and Fig. \ref{fig:Neutron}. (Right) Temperature dependent X-Ray diffraction of the compound NiMnCrAl over the range of 15~K to 300~K.}
	\label{fig:XRD}	
\end{figure*}

For a quaternary Heusler alloy of the type $XX^\prime YZ$, forming in Y-type ordered cubic structure, the structure factor for a general (hkl) reflection can be expressed as

\begin{equation}
	F_{hkl} = 4[f_{4a} + f_{4b} e^{\pi i(h+k+l)} + f_{4c} e^{\frac{\pi}{2i} (h+k+l)} + f_{4d} e^{ - \frac{\pi}{2i} (h+k+l)} ]
	\label{eqn:st1}
\end{equation}

\noindent
where $f_{4a}$, $f_{4b}$, $f_{4c}$, and $f_{4d}$ are the scattering factors of atoms at the $4a$, $4b$, $4c$ and $4d$ positions, respectively \cite{Rh2FeAlSudip}. For the first few Bragg peaks, the respective structure factors can then be written as

\begin{equation}
	\begin{split}
		F_{111} = 4[(f_{4a} - f_{4b}) - i(f_{4c} - f_{4d})]\\
		F_{200} = 4[(f_{4a} + f_{4b}) - (f_{4c} + f_{4d})]\\
		F_{220} = 4[(f_{4a} + f_{4b}) + (f_{4c} + f_{4d})]
	\end{split}
	\label{eqn:st2}
\end{equation}

\noindent
it is clear from Eq.~\ref{eqn:st2} that the intensities of the above-mentioned Bragg peaks depend not only on the distribution of the atoms in each Wyckoff position, but also on the scattering factors of individual constituent atoms \cite{XRDDRani}. Consequently, the diffraction pattern can help us not only to distinguish between Type-I, Type-II and Type-III atomic arrangements, but is also sensitive to the cross-site mixing of atoms that is often found in many Heusler alloys.  The extent of such atomic disorder can also be quantified using the ordering parameters $S$ and $\alpha$, defined as $S = \left[\frac{(I_{200}/I_{220})_{\mathrm{exp}}}{(I_{200}/I_{220})_{\mathrm{theory}}}\right]^{1/2}$ and $S(1-2\alpha) = \left[\frac{(I_{111}/I_{220})_{\mathrm{exp}}}{(I_{111}/I_{220})_{\mathrm{theory}}}\right]^{1/2}$ : fully ordered cubic phase is characterized by $S = 1$ and $\alpha = 0$; $L2_1$ \& $B2$-type disorder corresponds \cite{websterNDSalpha} to $S = 1$ and $\alpha = 0.5$ while for a complete A2-type disorder $S = 0$ and $\alpha = 0$. However, complexities may arise when the scattering factors of two or more atoms becomes comparable, where the above-mentioned analysis of diffraction pattern fail to produce any unique solution to crystal structure. X-ray diffraction (XRD) patterns of compounds containing elements with closely spaced atomic numbers get particularly vulnerable as the x-ray scattering cross-sections of those elements are almost indistinguishable.

The XRD pattern for NiMnCrAl taken at room temperature is presented in Fig.~\ref{fig:XRD}. Rietveld refinement confirms that all observed Bragg peaks can be indexed within the space group $F\bar{4}3m$ (No.~216), thereby establishing the single-phase nature of the synthesized compound. The lattice parameter is estimated to be 5.791(7)~\AA, in close agreement with values reported earlier \cite{CrNiMnAl-Published1,NiCrMnAl-Published2}. Following Eq. ~\ref{eqn:st2}, the absence of (111) Bragg peak intensity in our XRD data (Fig.~\ref{fig:XRD}) indicates that the scattering factors of atoms occupying the $4a$ and $4b$ as well as $4c$ and $4d$ sites are almost equal. However, our simulation of XRD patterns by considering all the three types of atomic arrangements (Table \ref{tab:magnetic_moments}) always result finite intensity for the (111) reflection, indicating the presence of disorder in the system. We found that the XRD pattern can be quantitatively explained almost equally well, when a complete disorder between only $4a$ (Al) and $4b$ (Mn or Cr or Ni) sites are considered ($L2_1$-type). Since the atomic number of Mn, Cr and Ni are very close, their respective x-ray scattering factors are also very similar and hence the XRD remain insensitive to the disorder between $4c$ and $4d$ sites and/or between $4c$/$4d$ and $4b$ sites. Neutron diffraction measurements, presented later, are capable of lifting such ambiguities. Fig. ~\ref{fig:XRD} presents the full rietveld refinement of XRD data using disorder between Al and Mn atoms within the Type-I atomic configuration. It may be noted here that similar disorder between $4a$ (Al) and $4b$ (Cr) sites in NiMnCrAl has also been reported earlier, who had proposed Type-II atomic arrangement (Table-\ref{tab:magnetic_moments}) to explain the experimentally obtained XRD pattern, although our theoretical investigation suggested that the stable minimum energy configuration can only be achieved by Type-I structure. On the other hand, The XRD pattern of same material reported in another work did not find both the (111) as well (200) Bragg peaks, implying complete atomic disorder among all 4 Wyckoff positions, as indicated by Eq.~\ref{eqn:st2}. To check the structural integrity of the sample over the studied temperature range, temperature-dependent powder X-ray diffraction patterns were recorded from 15 to 300~K [Fig.\ref{fig:XRD}]. No additional reflections, peak splitting, or noticeable peak broadening are observed down to the lowest temperature, indicating that the cubic structure is retained and that no structural phase transition occurs within the resolution of our measurements. A more detailed structural investigation through neutron diffraction (ND) and Extended X-ray Absorption Fine Structure (EXAFS) measurements have been presented later in section~\ref{sec:Neutron} and section~\ref{sec:EXAFS}, respectively.

\subsection{Theory: Effect of different atomic disorders}

\begin{figure*}
	\centering
	\includegraphics[width=16cm]{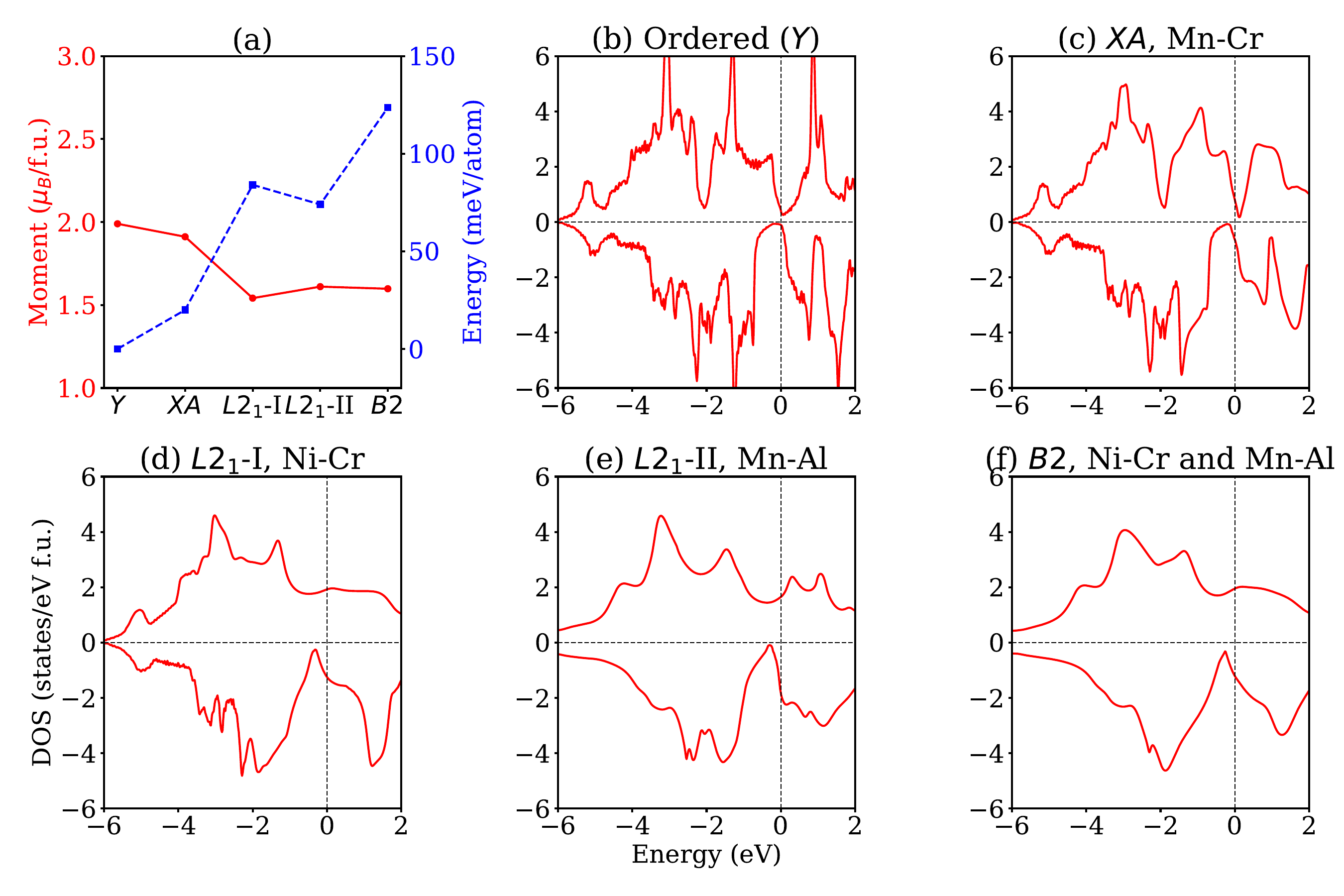}
	\caption{(a) Total magnetic moment (red, left axis) and relative total energy (blue dashed, right axis) of NiMnCrAl for the ordered $Y$-type structure and the disordered configurations $XA$, $L2_1$-I, $L2_1$-II, and $B2$. The energy of the $Y$-type structure is taken as the reference. Spin-resolved total density of states (DOS) for (b) the ordered $Y$-type structure, (c) $XA$-type disorder (Mn--Cr swap), (d) $L2_1$-I-type disorder (Ni--Cr swap), (e) $L2_1$-II-type disorder (Mn--Al swap), and (f) $B2$-type disorder (Ni--Cr and Mn--Al swaps). Positive and negative DOS values correspond to the spin-up and spin-down channels, respectively, and the Fermi level is set at zero energy (vertical dashed line).}
	\label{fig:DOSAll}
\end{figure*}

As mentioned earlier, due to the similar atomic sizes of a number of constituent atoms, many Heusler alloys are prone to internal elemental cross-substitution. Depending on the nature of disorder, the structure of quaternary Heusler alloys may further be sub-categorized as (i) $L2_1$-type, where intermixing is restricted only to one of the similar symmetric sites \textit{viz.}, either $4c$-$4d$ or $4a$-$4b$ sites, (ii) XA-type, where the intermixing involves only one of the octahedral $4a$ or $4b$ with one tetrahedral $4c$ or $4d$ sites, (iii) $B2$-type, where atomic disorder involves among the same symmetry positions only, \textit{viz.} between $4a$ \& $4b$ and $4c$ \& $4d$ sites, (iv) A2-type, where all the atoms display equal probability in occupying all the four Wyckoff positions \cite{T Graph}. Of course, strictly speaking, all these atomic disorders result in modification of site-symmetries, causing change in respective space-groups, but those can still roughly be considered as derivatives of the original structure of quaternary Heusler alloys (Space Group $F\bar{4}3m$). Our XRD analysis presented above suggests a complete random swapping between Mn and Al atoms leading to $L2_1$ disorder of the alloy. It is thus imperative to investigate the modification of magnetic properties for such disorder of NiMnCrAl vis-\`{a}-vis that for the ordered crystal structure estimated in section~\ref{sec:XRD}. Since the XRD analysis is insensitive to the disorder between Ni, Mn and Cr, and these disorders too, if present, would also affect the magnetic properties. Accordingly, we took this opportunity to theoretically study the possible effect of all such disorders, \textit{viz.}, (a) $L2_1$-I, resulting from complete swapping between Ni and Cr atoms, (b) $L2_1$-II, resulting from complete swapping between Mn and Al atoms, (c) $XA$, resulting from complete swapping between Mn and Cr atoms, and (d) $B$2, swapping between Ni and Cr atoms, and between Mn and Al atoms, simultaneously. Fig.-\ref{fig:DOSAll}(a) compares the stability of these disordered phases with respect to the ordered $Y$ phase. Magnetic moment found to be slightly reduced in $XA$-type to 1.91 $\mu_\mathrm{B}$ from the value of 1.99 $\mu_\mathrm{B}$ in the ordered $Y$-type. However, for $L2_1$-I, $L2_1$-II, and $B$2 types the magnetic moment is further reduced to 1.54 $\mu_\mathrm{B}$, 1.61 $\mu_\mathrm{B}$, and 1.60 $\mu_\mathrm{B}$, respectively. This reduction of the magnetic moment from its integer value of 2 $\mu_\mathrm{B}$ can be well understood from the DOS for different types of phases as a result of the swapping disorders. Fig.~\ref{fig:DOSAll} (b), (c), (d), (e), (f) represent the DOS for $Y$, $XA$, $L2_1$-I, $L2_1$-II, and $B$2 respectively. The minority DOS at the Fermi level are 0.65 states/eV. f.u., 1.26 states/eV. f.u., 1.90 states/eV. f.u., 1.22 states/eV. f.u. for $XA$, $L2_1$-I, $L2_1$-II, and $B$2 respectively. Thus, a minute density of states (DOS) in the minority-spin channel for the $Y$ and $XA$ types results in a near-integer magnetic moment in the NiMnCrAl alloy. Notably, despite this marginal reduction of moment, the $XA$-type disorder does not destroy the SGS character, as the minority-spin DOS at $E_\mathrm{F}$ remains vanishingly small, preserving the gap-like feature in that channel. However, the same Mn-Cr site-mixing that underlies the $XA$-type disorder also perturbs the dominant Mn-Cr exchange pathway, thereby weakening the net exchange interaction strength and yielding a slightly reduced transition temperature ($T_ \mathrm C$) relative to that of the ordered $Y$-type phase. One may note here that among all the different types of disorders discussed here, the $L2_1$-disorder in NiMnCrAl could be unambiguously detected through XRD analysis, which would be unresponsive to $B2$-disorder, if present in the system. Incidentally, both the disorder types result in very similar reduced moment of $\sim$1.6 $\mu_\mathrm{B}$/f.u.

\subsection{DC Magnetism}

\subsubsection{Temperature-dependent magnetisation}

\begin{figure}[ht]
	\centering
	\includegraphics[width=8cm]{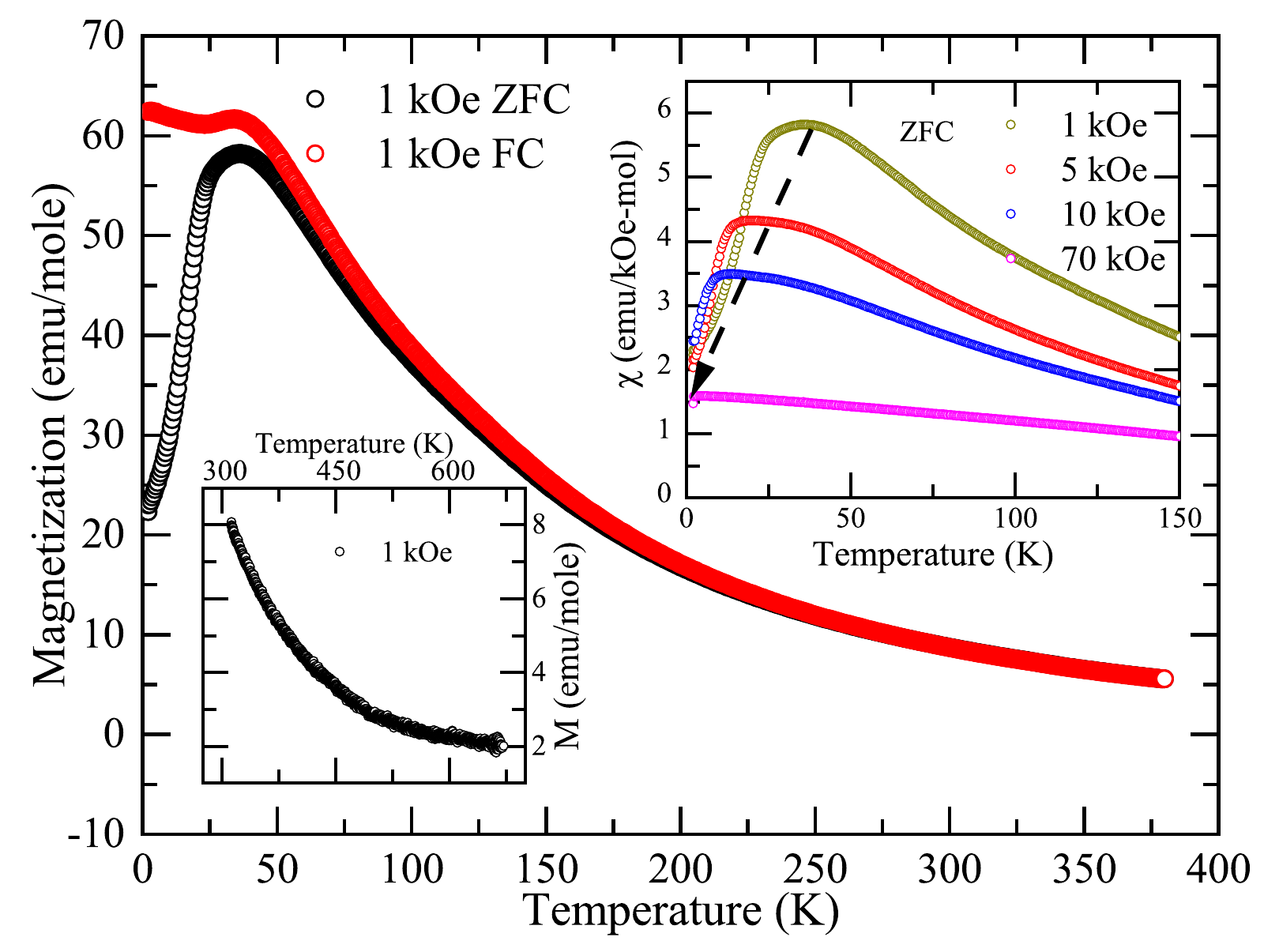}
	\caption{magnetisation with Temperature, M(T), under 1 kOe applied external field in both Field-Cooled (FC) and Zero-Field-Cooled (ZFC) protocol. In top inset, the field dependence of the magnetic susceptibility, $\chi$, is shown. In the bottom inset, the temperature dependence of magnetisation from 300 K to 700 K is shown. }
	\label{fig:MT}
\end{figure}

As mentioned earlier, our \textit{ab-initio} calculations on the ordered $Y$-type phase of NiMnCrAl indicated a magnetic phase transition near $\sim$1300~K whereas, the disorder-driven suppression should reduce the transition temperature to 500-600~K for the $B2$ or $L2_1$ phases. Accordingly, we first investigated the magnetic susceptibility behaviour, M(T), in the temperature range above the room temperature. Due to the limitation of our high temperature VSM set-up, our measurement is restricted in the temperature region 300-700~K. The measurement was carried out under a fixed applied field of 1000~Oe. The experimental magnetisation decreases monotonically with increasing temperature, similar to that found in a paramagnetic material, indicating the absence of any magnetic phase transition for T $>$ 300~K (Inset I: Fig.~\ref{fig:MT}). Subsequently, we have extended the measurement in low temperature region, 5 $\leq$ T $\leq$ 390~K, again under a field of 1000~Oe.  The measurements were performed under both the Zero-Field-Cooled Warming (ZFC) and Field-Cooled Warming (FC) \cite{RuMnGaSudip} protocols (Fig.~\ref{fig:MT}). In the ZFC protocol, the sample was first cooled to 5~K in the absence of any applied magnetic field. Once the base temperature was stabilized, magnetisation was recorded as a function of increasing temperature up to 390~K. The FC protocol was performed similarly, except that the magnetic field was applied during cooling. Data were then recorded upon warming, once the base temperature stabilized.

Both the ZFC and FC measurements indicated a broad hump in M(T) around $\sim$37~K, accompanied by a clear bifurcation in magnetisation measured under ZFC and FC protocols. The bifurcation of susceptibility under ZFC and FC measurements start to manifest at a much higher temperature, $\sim$150~K. A similar result was earlier reported for this compound, and the anomaly at $\sim$37~K had been explained there as compensated ferrimagnetic phase transition. We have further extended our M(T) measurements under various external magnetic fields (Inset II: Fig.~\ref{fig:MT}) and found that the peak position shifts towards lower temperature as the field strength is increasing. Such characteristic resembles more with antiferromagnetic spin alignments instead of ferrimagnetic phase transition reported elsewhere \cite{CrNiMnAl-Published1}. Of course, on a closer look, the susceptibility peak appears to be much broader than standard antiferromagnetic systems. Together with the pronounced FC-ZFC bifurcation, the dc magnetic susceptibility resemble more closely to metastable spin arrangements, rather than any long-range magnetic ordering.

\subsubsection{Isothermal magnetisation}

\begin{figure}
	\centering
	\includegraphics[width=8cm]{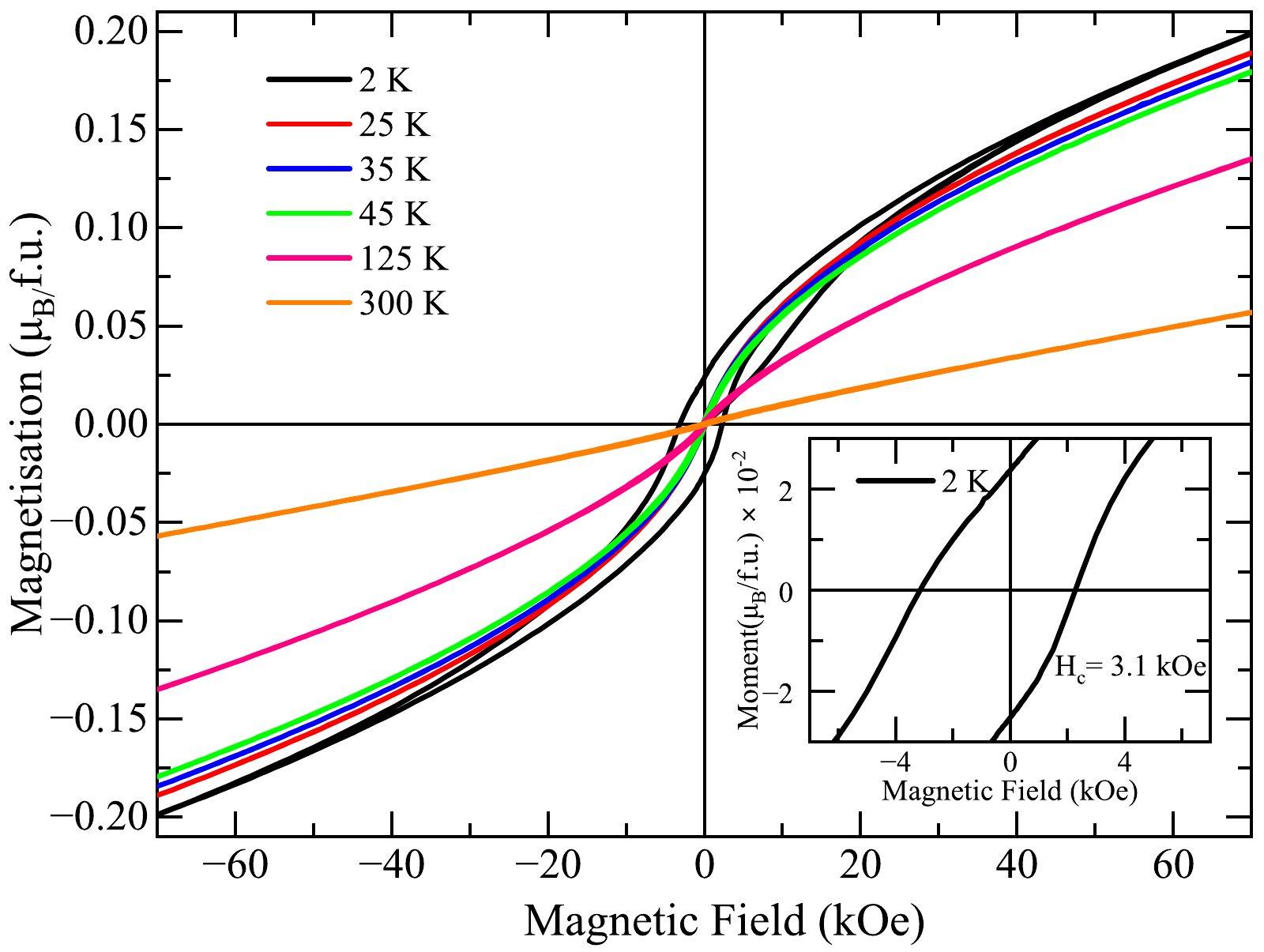}
	\caption{magnetisation (M) with applied external field (H) in various temperature range from 2~K to 300~K. Inset shows the enlarged version of 2~K mgnetisation.}
	\label{fig:MH}	
\end{figure}

Isothermal magnetisation mesurements, M(H), at different temperatures were carried out to further probe the magnetic properties. Fig.~\ref{fig:MH} presents the M(H) over the magnetic field range of $\pm$70~kOe at different fixed temperatures, \textit{viz.}, 2~K, 25~K, 35~K, 45~K, 125~K, and 300~K. At 300~K, the magnetisation varies almost linearly with the applied field over all the magnetic field range used in this measurement, substantiating the paramagnetic state at room temperature indicated earlier by the magnetic susceptibility results. As the temperature is lowered to 125~K, the isotherm begins to acquire a discernible curvature at low fields, indicating the onset of short range magnetic correlations, although the response still remains predominantly linear at higher fields. The weak nonlinear behaviour of isothermal magnetisation at 125~K is consistent with the magnetic susceptibility results showing a bifurfaction between ZFC and FC measurements below $\sim$150~K. On further cooling to 45~K, 35~K, and 25~K, the isotherms display a pronounced sigmoidal shape with rapid growth of the magnetisation at low fields followed by a much slower, quasi linear increase at higher fields. The lowest temperature isotherm, recorded at 2~K, shows the steepest initial rise among all the curves and also exhibits a small but clearly resolved hysteresis of $H_c = 3.1$~kOe (Fig. \ref{fig:MH} inset). Notably, none of the isotherms, including the one recorded at 2~K, attain complete saturation up to the maximum field of 70~kOe, with the measured moment reaching only a value of about 0.20~$\mu_\mathrm{B}$/f.u. at this field and temperature.

The low value of saturation moment, M$_S$, observed at the lowest temperature, 2~K, is indeed surprising, as our theoretical calculations predict an expected value of M$_S$ to be approximately $2$~$\mu_ \mathrm B$/f.u. for NiMnCrAl in both the ordered Type-I and XA type structures which are expected to show the spin-gapless semiconducting behavior that corroborates the Slater–Pauling rule. Even in case of disordered $L2_1$ type structure, that looses the SGS feature yet maintaining the long range ferromagnetic ordering, the predicted magnetic moment is reduced to approximately $1.6$~$\mu_ \mathrm B$/f.u. only. Furthermore, we find that the magnetisation does not attain saturation even at the highest applied field, and this `S'-shaped nature of the M(H) curve is clearly inconsistent with either ferromagnetic or ferrimagnetic character. Additionally, although the shift of peak position in the M(T) curve as a function of applied magnetic field might be suggestive of an antiferromagnetic transition, the antiferromagnetic systems typically exhibit neither sigmoidal magnetisation behavior, nor display hysteresis. Considering all these together, one can infer that the magnetic anomaly at $\sim$37~K cannot be attributed to any conventional long-range ferro/ferri/antiferro-magnetic phase transition, but point towards the presence of a distinct and complex metastable magnetic state.

\subsubsection{Magnetic Relaxation}

\begin{figure}
	\centering
	\includegraphics[width=8cm]{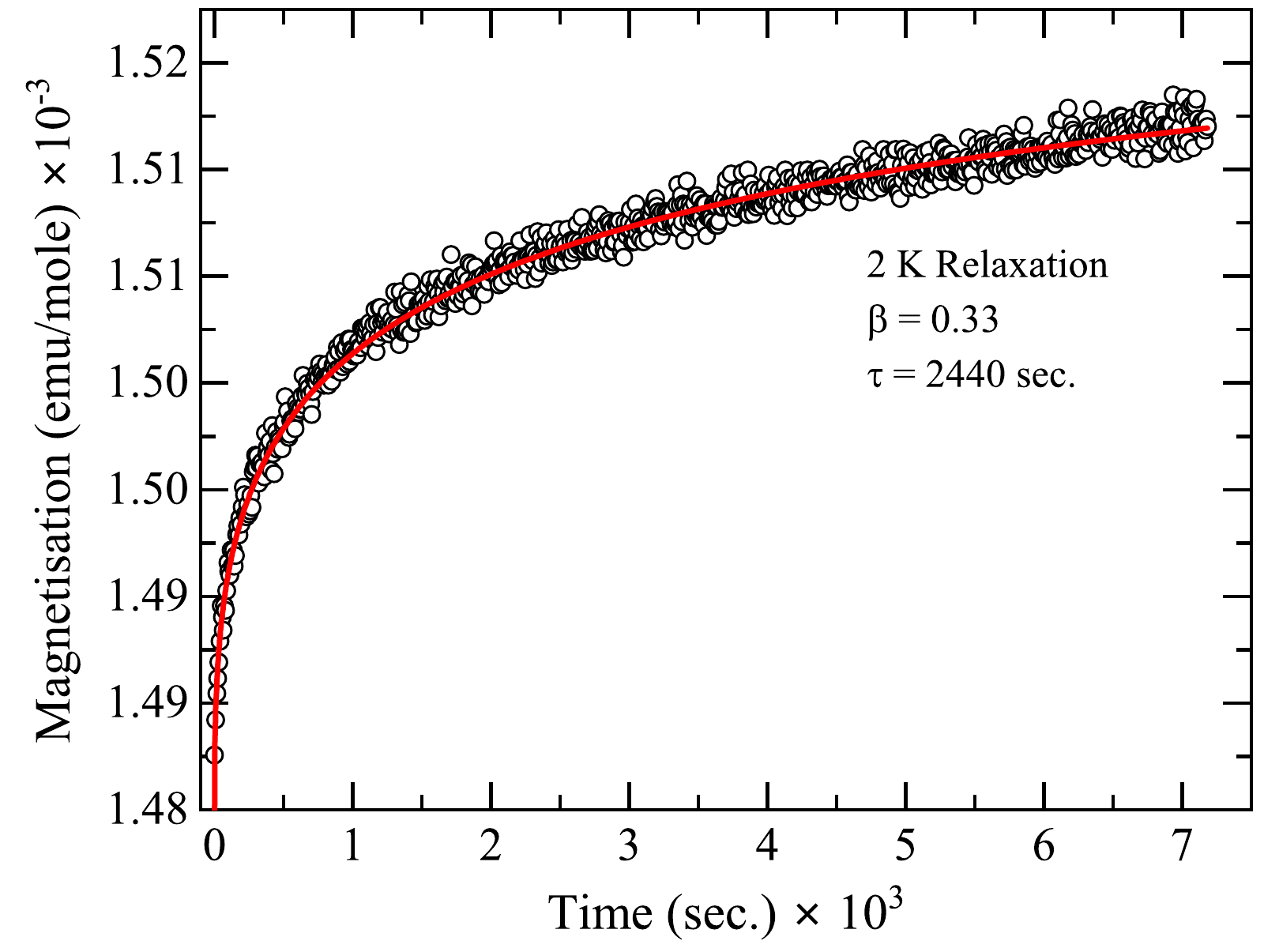}
	\caption{Magnetic relaxation measured in Zero-Field-Cooled protocol (ZFC) at temperature T = 2 K for the wait time t$_w$ of 7500 sec. The red line represents fit to the data using KWW equation, mentioned in Eqn.~\ref{eqn:KWW}.}
	\label{fig:Relaxation}
\end{figure}

To check the presence of metastable spin state in NiMnCrAl, we have carried out standard Zero-Field-Cooled (ZFC) magnetic relaxation measurements \cite{milydipr}. In this protocol, the sample was first cooled from the paramagnetic state (300~K) to low temperature (2~K) in the absence of any external magnetic field. After stabilizing the system at 5~K by waiting for 60~sec., a field of 100 Oe was applied, and the time-dependent magnetisation, M(t), was recorded for a duration of 7500 sec.

The relaxation data thus obtained (Fig.~\ref{fig:Relaxation}) exhibited an exponential increase in magnetisation with time, characteristic of multiple competing magnetic spin-states, such as spin-glasses or superparamagnets. To quantitatively analyze the relaxation behaviour, the experimental data were fitted using the Kohlrausch–Williams–Watts (KWW) stretched exponential function \cite{mypaper}:

\begin{equation}
	\label{eqn:KWW}
	M(t) = M_0-M_r \exp\left[-\left(\frac{t}{\tau}\right)^{\beta}\right]
\end{equation}

\noindent
here, $M_0$ is the initial magnetisation and $M_r$ is the relaxing component of the magnetisation. While $\beta = 1$ corresponds to a single relaxation time, a non-integer value, $0<\beta<1$ indicates a broad distribution of relaxation times, arising from multiple local energy minima within the spin-configuration. For a typical magnetically glassy phase with a wide barrier distribution, $\beta$ is expected to lie in the range of 0.2 to 0.6 \cite{mypaper}. We find that relaxation phenomenon of NiMnCrAl can be fitted well with $\beta \sim 0.33$ confirming the relaxation process being governed by a wide distribution of energy barriers, consistent with the frustrated nature of the spin-configuration. The metastable states in NiMnCrAl results in a slow, time-dependent evolution towards the equilibrium under applied magnetic field.

\subsubsection{Magnetic memory}

\begin{figure}
	\centering
	\includegraphics[width=8cm]{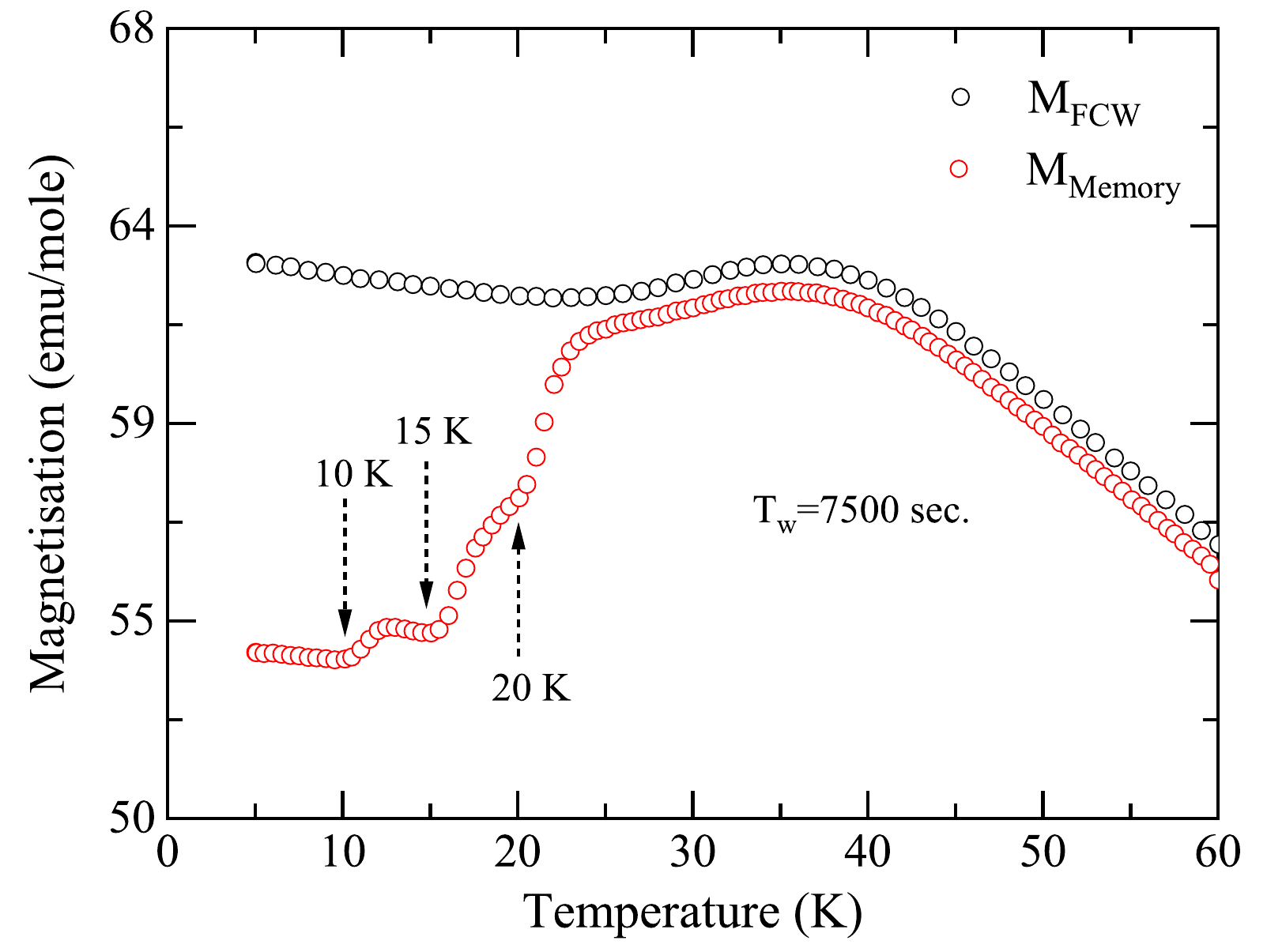}
	\caption{$\mathrm M_{\mathrm{FCW}}$ ($\circ$) is the field-cooled warming reference curve, while $\mathrm M_{\mathrm{Memory}}$ (\textcolor{red}{$\circ$}) was recorded on warming after intermittent halts of $\mathrm T_ \mathrm w = 7500$~s during cooling at the temperatures marked by the three dotted arrows (10, 15 and 20~K). $\mathrm M_{\mathrm{Memory}}$ shows step-like anomalies at each halt temperature.}
	\label{fig:FC Memory}
\end{figure}

Magnetic memory is another tool that can be used to probe the metastable spins with quite certainty. Similar to many other magnetic measurements, the memory effect are also measured under both the FC and ZFC protocols \cite{santanuda213, SunPRL}. For the FC protocol, the sample was cooled from 300 K to 5 K under a constant magnetic field of 100 Oe. During the cooling, stops of 7500 sec. each were introduced at 20 K, 15 K, and 10 K (Fig.~\ref{fig:FC Memory}). The magnetisation was recorded continuously during warming up to 60 K under the applied field. Upon warming, distinct anomalies or steps were observed at each halt temperature (10 K, 15 K, and 20 K), demonstrating the presence of magnetic memory effects at all the three stop-temperatures. The system thus retains imprints of its thermal and magnetic history experienced during these stops, confirming the presence of metastable spins. However, it may be noted that the FC memory effect can not distinguish between the superparamagnetic and spin/cluster-glass systems, as metastable magnetic ground state is found in both of them. However, unlike FC memory effect, ZFC memory effect is unique to spin/cluster glasses.

\begin{figure}
	\centering
	\includegraphics[width=8cm]{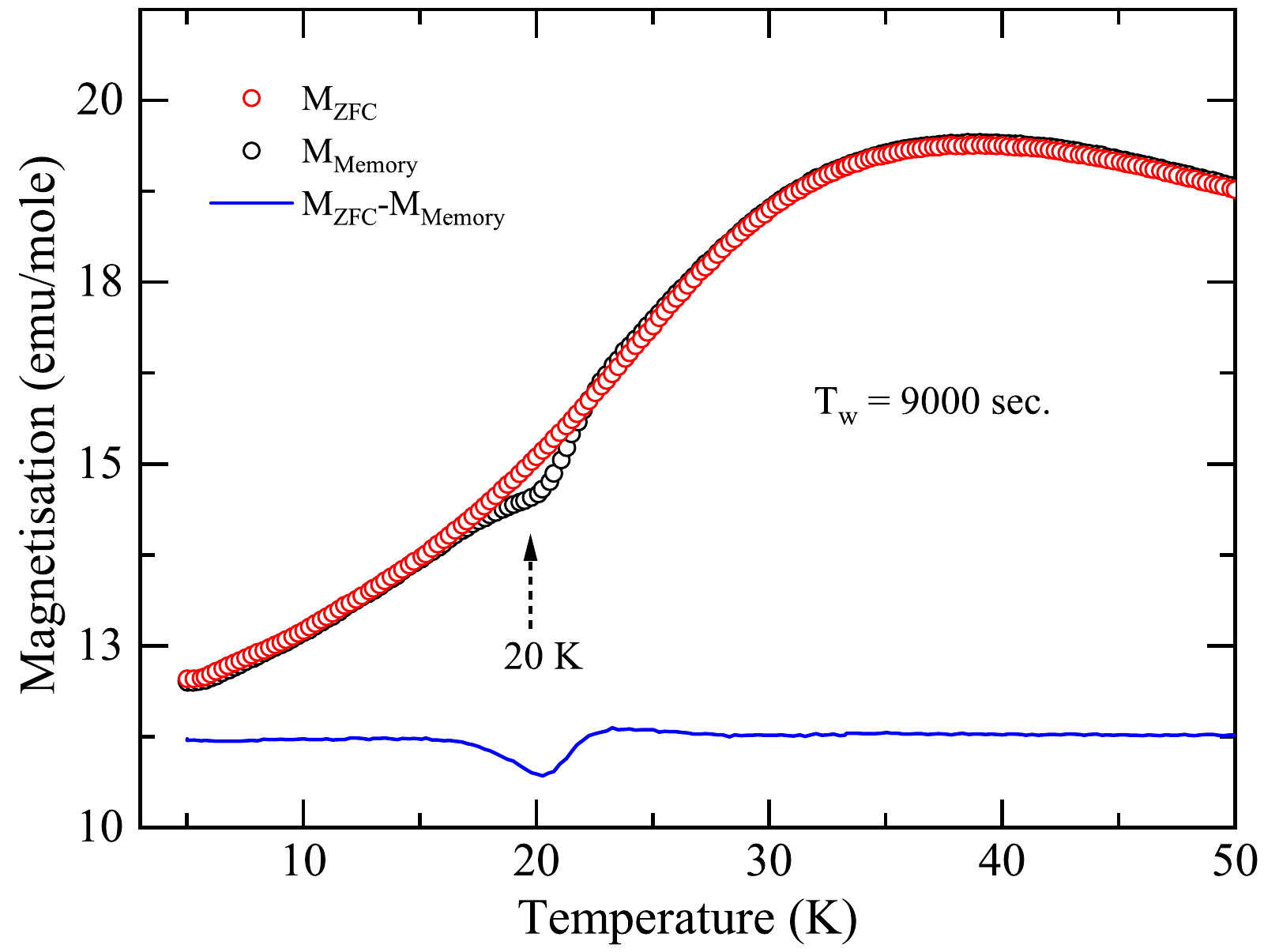}
	\caption{Zero-Field-Cooled (ZFC) memory with a waiting time of 9000 sec. at 20 K during cooling.}
	\label{fig:ZFC Memory}
\end{figure}

For the ZFC protocol, the sample was cooled from 300 K to 5 K in the absence of any external magnetic field. During cooling, a halt was introduced at 20 K for a waiting time of 9000 sec. to allow partial equilibration of the spin configurations. After reaching 5 K, a magnetic field of 100 Oe was applied, and the magnetisation, M(T), was recorded during warming from 5~K to 50~K (Fig.~\ref{fig:ZFC Memory}). A clear dip was observed at 20 K, corresponding to the halt temperature, indicating the magnetic memory effect arising from the freezing of spin configurations during the wait-period. This demonstrates that the system retains memory of its thermal history due to its non-ergodic, metastable nature. Observation of ZFC memory effect suggests the spin/cluster glass phase, instead of superparamagnetic spin-state.

\begin{figure*}
	\centering
	\includegraphics[width=18.5cm]{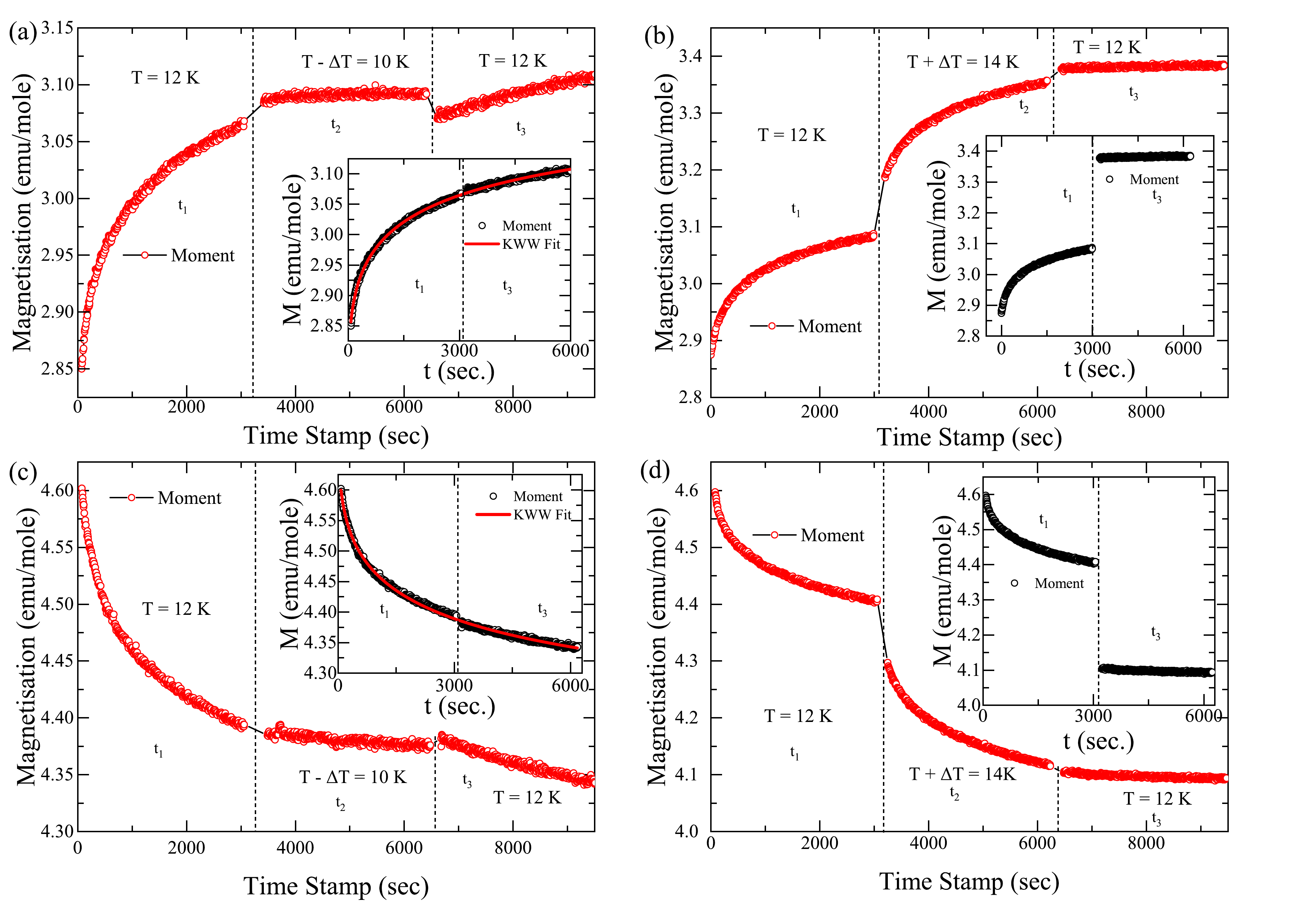}
	\caption{Time dependent magnetisation at $T = 12$~K following a negative [(a), (c), $T - \Delta T = 10$~K] and a positive [(b), (d), $T + \Delta T = 14$~K] temperature-cycling protocol. The dip/step observed at $t_2$, followed by recovery of the original $t_1$ relaxation trend in $t_3$, confirms the memory effect characteristic of spin-glass dynamics. Insets: the $t_1$--$t_3$ relaxation data fitted with the KWW equation.}
	\label{fig:RM}
\end{figure*}

The memory effect of spin/cluster glass systems can be understood with different theoretical models, depending on the nature of their time-dependent magnetisation response. At a given temperature, the magnetic system may possess a single, favored spin configuration. Such behavior is described within the framework of droplet model \cite{droplet}. Alternatively, the system may be governed by a multi-valley free-energy landscape, comprising numerous nearly degenerate metastable states separated by barriers of varying height. This scenario is generally associated with the hierarchical model \cite{hierarchical1, hierarchical2}. These two models can be experimentally distinguished by checking whether the original spin configuration is recovered after a positive temperature cycle: in the droplet model the original configuration is restored, whereas in the hierarchical model, temporary heating rejuvenates the system and erases the original configuration. To probe which mechanism governs the memory effect in NiMnCrAl, appropriate measurements under two temperature cycling were carried out under both the FC and ZFC condition (Fig.~\ref{fig:RM}), following a previously reported protocol \cite{santanuda213,SunPRL}.

For the ZFC relaxation measurement, the sample was cooled in zero-field to $T_0 = 12$~K ($T_0 < T_f$, where $T_f \approx 37$~K). A field of 100~Oe was then applied, and the magnetisation M(t) was recorded for $t_1 = 3000$~sec. The sample was subsequently quenched to a lower temperature, $T_0 - \Delta T = 10$~K, at the same field, and $M(t)$ was recorded for $t_2 = 3000$~sec. Finally, the temperature was raised back to $T_0 = 12$~K, and $M(t)$ was recorded for a further period of $t_3 = 3000$~sec. In the FC protocol, the sample was instead cooled to $T_0$ under the applied field of 100~Oe, and the same subsequent sequence was followed after switching off the field.

The outcome of the memory measurement, thus obtained, is presented in Fig.~\ref{fig:RM} (a) and Fig.~\ref{fig:RM} (c). The results in both the negative temperature cycles show that the magnetisation recovers to the level it had reached prior to the temporary cooling once the temperature is brought back to $T_0$, and the relaxation curve during $t_3$ appears as a smooth continuation to that during $t_1$. Both segments could be fitted simultaneously with a single stretched-exponential function as explained in Eq.~\ref{eqn:KWW} yielding $\beta \approx 0.31$ which is consistent with typical magnetic-glass relaxation behavior.

The effects of positive temperature cycling were also examined. Here, the sample was cooled to $T_0 = 12$~K in zero-field, then a field of 100~Oe was applied, and $M(t)$ was recorded for $t_1 = 3000$~sec. The sample was then heated to a higher temperature, $T_0 + \Delta T = 14$~K, at constant field, and M(t) was recorded for $t_2 = 3000$~sec. Finally, the temperature was brought back down to $T_0 = 12$~K, and M(t) was recorded for $t_3 = 3000$~sec. Unlike in the negative cycling cases, the magnetisation at the start of $t_3$ does not recover to its pre-heating level, and the relaxation behavior during $t_3$ differs markedly from that during $t_1$ (Fig.~\ref{fig:RM} (b) \& Fig.~\ref{fig:RM} (d)). This indicates positive temperature cycling rejuvenates the relaxation process, and hence erases the memory effect.

The observed memory effects in both ZFC and FC protocols can be directly attributed to the spin/cluster glass system, where halts allow the system to age and explore deeper local energy minima, resulting in the freezing of metastable spin configurations. These results provide a clear evidence of non-equilibrium spin dynamics and aging behavior which are typical for spin/cluster glass. This asymmetric response to negative versus positive temperature cycling memory preserved on cooling but erased on heating favors the hierarchical model over the droplet model for describing the memory behavior in NiMnCrAl. Together, these magnetic relaxation and memory-effect studies confirm the glassy magnetic phase in this compound.

\subsection{AC Magnetism}

\begin{figure}
	\centering
	\includegraphics[width=8cm]{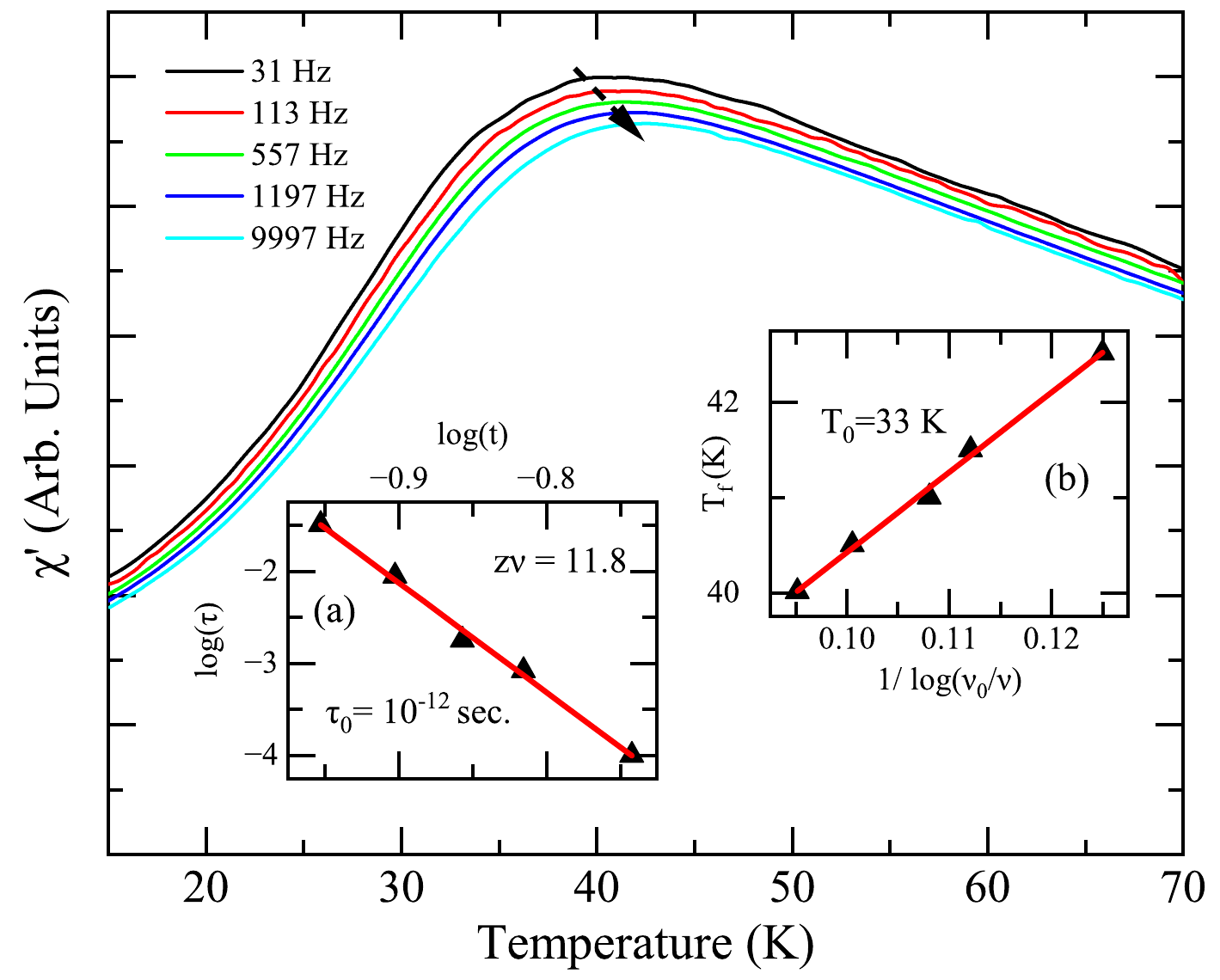}
	\caption{Temperature dependence of the real-part of magnetic ac-susceptibility over the range of 15~K to 70~K. In the inset (a) the power law analysis and (b) the Vogel-Fulcher and is presented.}
	\label{fig:AC}
\end{figure}

Temperature-dependent ac susceptibility measurements at different frequencies are commonly used to establish the metastable nature of a magnetic material. For such a system, the peak position shifts towards higher temperatures with increasing frequency. Fig. \ref{fig:AC} shows the temperature variation of the real part of the ac susceptibility, $\chi^\prime$(T), measured under a low ac field of 10~Oe at different frequencies. The $\chi^\prime$(T) data exhibit a well-rounded peak near $\sim$ 37~K. The peak shifts with frequency, indicating that the anomaly originates from the magnetically glassy phase. To determine the nature of this spin-glass state - whether canonical or cluster-glass - a detailed analysis of the temperature variation of the $\chi^\prime$(T) peak was carried out.

As mentioned above, the freezing temperature ($T_f$) of a magnetic glass tends to shift towards higher temperatures with increasing frequency. The relative shift in the freezing temperature per decade of frequency, generally known as Mydosh parameter \cite{mydosh,Dy2IrSi3-Sudip}, can be formulated as

\begin{equation}
	\delta(T_f) = \frac{\Delta(T_f)}{T_f \Delta (log_{10}f)}
\end{equation}

\noindent
here $\Delta{T_f}$ is the change of freezing temperature with frequency, \textit{f}. The parameter $\delta{T_f}$ provides information about the type of spin-glass: for our system this value came to be 0.02. $\delta{T_f}$ usually takes a value in the order of 0.001 for a canonical glass and 0.01 for cluster-glass. Our estimation of $\delta{T_f} \approx$ 0.02 thus suggest our sample of interest lies in the zone of cluster-glass.

The conventional power law can also be utilized to understand the magnetically glassy phase \cite{chikazumi}. For this analysis, one uses the equation

\begin{equation}
	\tau = \tau_0 \left[\frac{T_f - T_{SG}}{T_{SG}} \right] ^{-{z\nu}}  	
\end{equation}

\noindent
here $\tau$ is the relaxation time associated with the frequency ($\tau$ = $\frac{1}{\textit{f}}$), $\tau_0$ is called the single flip relaxation time, $T_{SG}$ is the spin glass temperature found in the DC measurement with \textit{f} = 0 condition and $z\nu'$ denotes the dynamic critical exponent associated with the correlation length, defined as $\xi = \left(\dfrac{T_f}{T_{SG}} - 1\right)^{-\nu'}$. In this expression, the parameters $z\nu$ and $\tau_0$ provide the information about the nature of glassy phase. In the studied compound, the value of $z\nu$ is estimated to be 11.8. For typical spin-glass systems, $z\nu$ lies in the range of 4 to 12. The obtained value therefore falls well within this range supporting the formation of a spin-glass state in the compound. Concomitantly, the value of $\tau_0$ is determined to be $\sim 10^{-12}$~sec. Usually the value $\tau_0$ lies in range of $10^{-13}$ to $10^{-12}$ sec. for canonical spin-glass and $10^{-11}$ to $10^{-4}$ sec. range for cluster-glass \cite{mypaper,santanuda213,RuMnGaSudip}. This result places the compound within the borderline framework of the canonical and cluster-glass model.

Another method known as the Vogel-Fulcher Law can be used to determine the frequency-dependence of the spin-freezing temperature \cite{Kaczorowski,mypaper}. The Vogel-Fulcher law is defined as

\begin{equation}
	f = f_0 exp \left[\frac{-E_a}{ K_B (T_f - T_0)}\right]
\end{equation}

\noindent
in this equation, $E_a$ represents the activation energy reflecting the strength of interaction among the frustrated spins, $T_0$ is the Vogel-Fulcher temperature, $f_0$ denotes the characteristic frequency, and $k_B$ is the Boltzmann constant. For canonical spin glasses, the ratio $E_a/(k_B T_0)$ typically approaches unity. In cluster-glass systems, this ratio takes a comparatively higher value. For the present compound, $E_a/(k_B T_0)$ is found to be $\sim 2.5$. The $T_f$ vs $1/\log_{10}(\nu_0/\nu)$ plot for NiMnCrAl yields a fitted value of $T_0 \approx 33$~K. These value lies within the cluster-glass regime.

\subsection{Thermal Conductivity and Seebeck Coefficient}

\begin{figure}
	\centering
	\includegraphics[width=8cm]{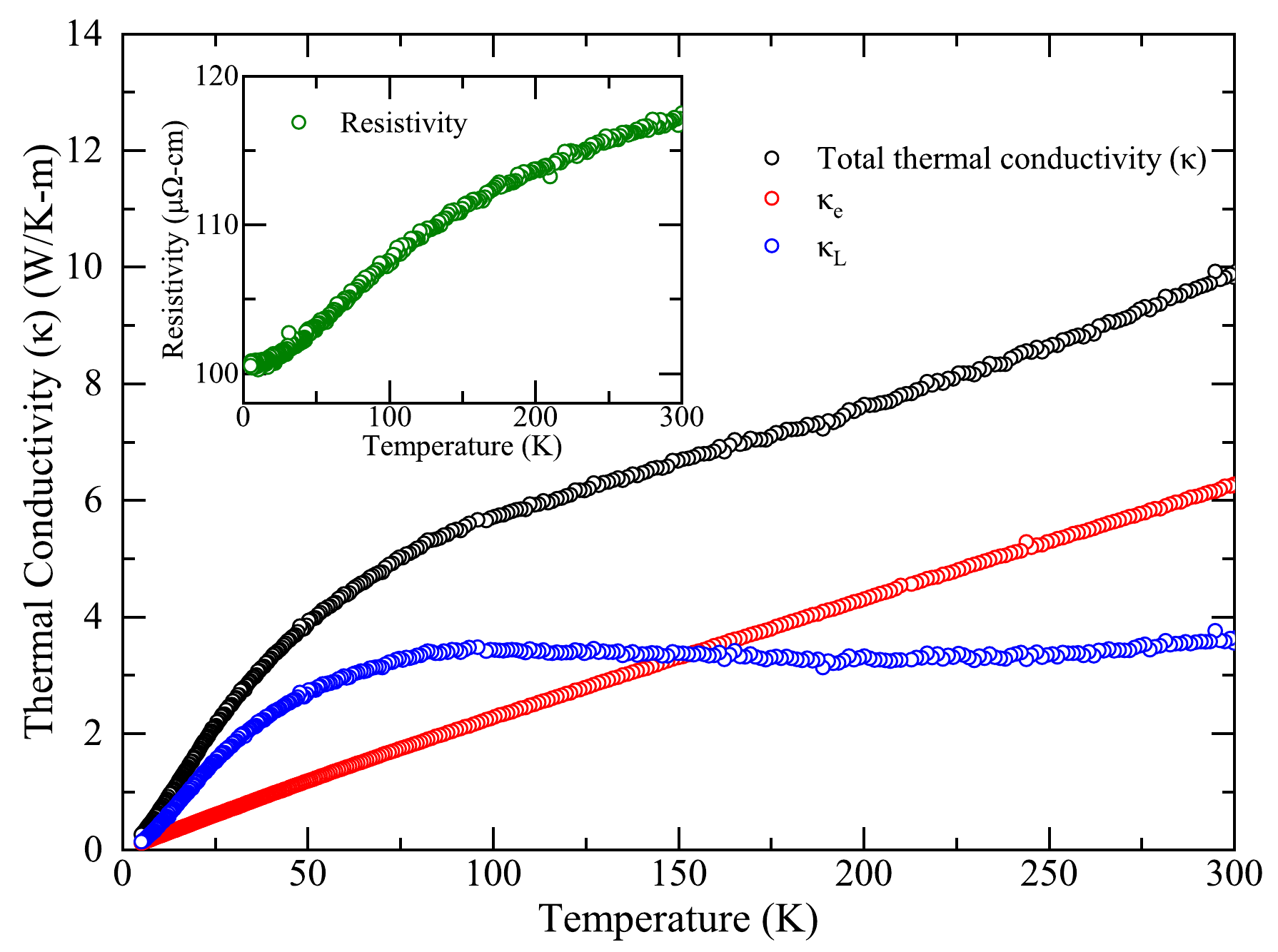}
	\caption{Temperature dependence of total ($\kappa$), electronic ($\kappa_e$), and lattice ($\kappa_L$) thermal conductivity. $\kappa_e$ is estimated using the Wiedemann-Franz law, and $\kappa_L$ is obtained by subtraction. Inset shows electrical resistivity, confirming metallic behavior. }
	\label{fig:Thermal Conductivity}
\end{figure}

\begin{figure}
	\centering
	\includegraphics[width=8cm]{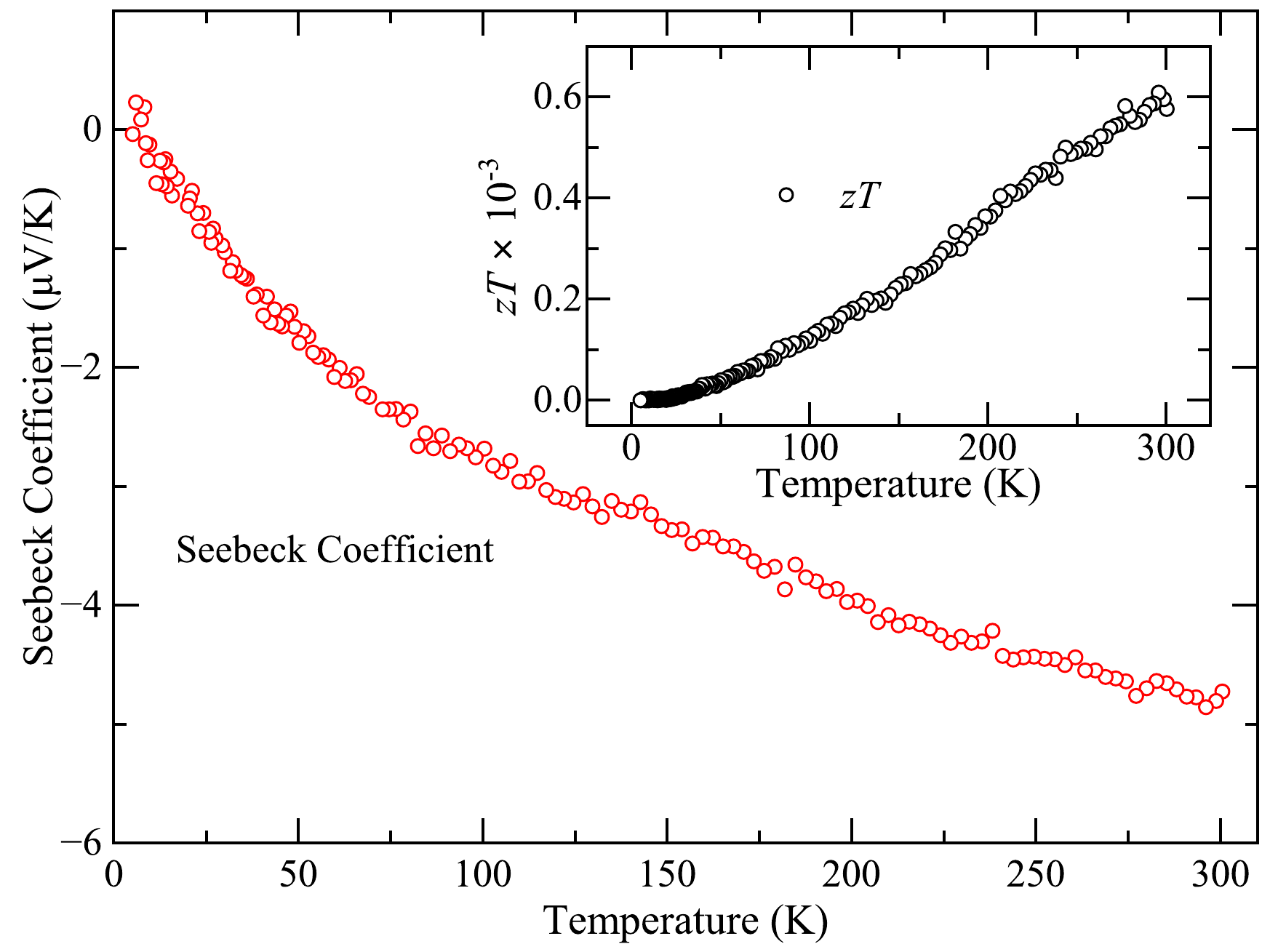}
	\caption{Seebeck Coefficient in the temperature range of 5~K to 300~K. Inset shows the figure of merit (\textit{ZT}) in the temperature range 5~k to 300~K.}
	\label{fig:Seebeck}
\end{figure}

To conclusively eliminate the possibility of retaining Spin-Gapless-Semiconducting feature, we have carried out electrical as well as thermal transport characterizations in the temperature range 5-300~K. The temperature dependencies of electrical resistivity $\rho(T)$ and thermal conductivity $\kappa(T)$ of NiMnCrAl are shown in Fig.~\ref{fig:Thermal Conductivity}, while the temperature dependence of the Seebeck coefficient $S(T)$ and thermoelectric figure of merit (\textit{zT}) are presented in Fig.~\ref{fig:Seebeck}. In NiMnCrAl, the overall value Seebeck coefficient is negative, which suggests the primary charge carriers to be electrons. This conclusion matches with the results obtained from the electrical resistivity measurement as the $\rho$$_{xx}(T)$ increases for increasing temperature showing metallic behavior (Fig~\ref{fig:Thermal Conductivity}(inset)). Mathematically, seebeck coefficient which defined as the ratio of thermoelectric voltage difference ($\Delta$V) and temperature difference at the junctions ($\Delta$T) and can be expressed by simplified Drude-Sommerfeld model:
\begin{equation}
	S(T) = -\frac{8\pi^2 k_B^2 T}{3eh^2}m^*
	\left(\frac{\pi}{3n}\right)^{2/3},
\end{equation}
where $e$ is the electronic charge, $n$ is the density of the charge carriers, $m^*$ is the effective mass, $k_B$ is the Boltzmann constant, and $h$ is Planck's constant \cite{seebeckformula}. The Seebeck coefficient $S(T)$ exhibits non-linear behavior at low temperatures, but appears to yield a more linear trend above $\sim$125~K. The Seebeck coefficient at 300~K is estimated to be -4.8~$\mu$V/K. This value is in good agreement with those reported for other Heusler alloys \cite{sanchayita, thermoelectric2}. The metallic resistivity behavior, together with electrons being the primary charge carriers, rules out the possibility of NiMnCrAl hosting SGS characteristics. Additionally, thermoelectric response of the present compound is investigated using thermal conductivity. The electronic thermal conductivity ($\kappa_e$) given by Wiedemann–Franz law is $\kappa _e \rho /T = \rm L _0$ where $\rho$ is the electrical resistivity and L$_0$ is the Lorrentz number given as L$_0$ = 2.45 $\times$ $10^{-8}$ W~$\Omega~\text{K}^{-2}$. Using the above expression $\kappa_e$ can be easily estimated from the resistivity data \cite{thermoelectric3}. Subtracting the electronic thermal conductivity ($\kappa_e$) from the total thermal conductivity ($\kappa$), we can get the Lattice thermal conductivity ($\kappa_L$) \cite{ThermalConductivity1,ThermalConductivity2}. Fig.~\ref{fig:Thermal Conductivity} shows a sharp increase in $\kappa$ up to $\sim$50~K, followed by a gradual increase up to 300~K. The thermal‑conductivity spectrum shows no distinct low‑temperature peak, a feature that typically emerges from the competition between defect‑driven phonon scattering at low temperatures and Umklapp scattering that becomes dominant at higher temperatures \cite{kittel}. Given that a sharp maximum in $\kappa(T)$ typically correlates with structural periodicity and atomic ordering in Heusler compounds, its absence can thus be linked to underlying atomic disorder \cite{sanchayita, thermoelectric2}. Using all three measured thermoelectric parameters, \textit{viz.}, the Seebeck coefficient $S(T)$, electrical resistivity $\rho(T)$ and thermal conductivity $\kappa(T)$, the dimensionless figure of merit $\left(zT = S^{2}T/\rho\kappa\right)$ is estimated and illustrated in Fig.~\ref{fig:Seebeck}(inset). It is still several orders of magnitude smaller than that of the state-of-the-art thermoelectric material Bi$_{2}$Te$_{3}$ at 300~K \cite{zTOne}, which displays $zT \approx 1$, but lies within the usual range for those of the Heusler alloys \cite{SanchayitadizT}.

The absence of a pronounced peak in $\kappa(T)$ points toward substantial local structural disorder in NiMnCrAl, a scenario independently corroborated by ac and dc magnetisation measurements, revealing a spin-glass-like ground state. Conversely, Rietveld refinement of XRD patterns yields a long-range $L2_1$ ordered structure with a saturation magnetic moment of $\sim$1.6~$\mu_\mathrm{B}$/f.u. To reconcile this apparent dichotomy between long-range periodic symmetry and local disorder, extended X-ray absorption fine structure (EXAFS) measurements were conducted to probe the short-range atomic environment directly.

\subsection{EXAFS}
\label{sec:EXAFS}

\begin{figure*}
	\centering
	\includegraphics[width=16cm]{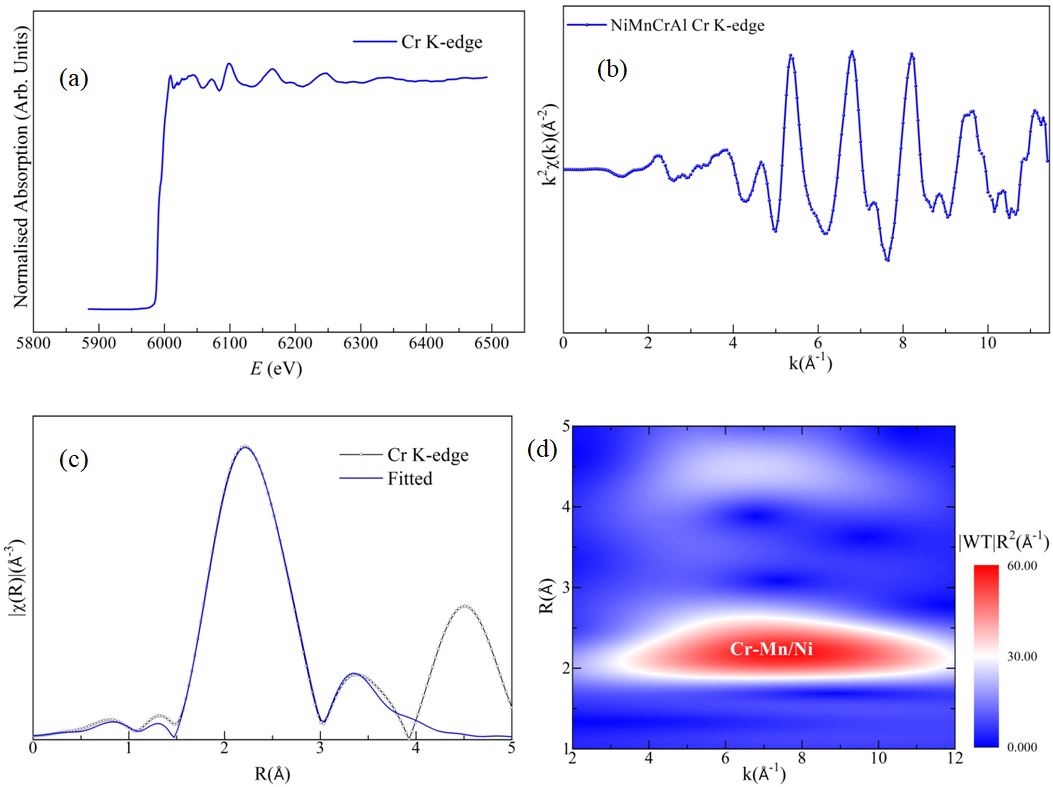}
	\caption{(a) Normalized Cr $K$-edge X-ray absorption spectrum of NiMnCrAl. (b) Background-subtracted EXAFS oscillations, $k^{2}\chi(k)$, at the Cr $K$-edge. (c) Fourier-transformed magnitude $|\chi(R)|$ of the Cr $K$-edge EXAFS data (open circles) along with the best fit (solid line). (d) Wavelet-transform (WT) EXAFS contour plot of the Cr $K$-edge, showing the dominant contribution from the Cr--Mn/Ni scattering path.}
	\label{fig:EXAFS-Cr}
\end{figure*}

\begin{figure*}
	\centering
	\includegraphics[width=16cm]{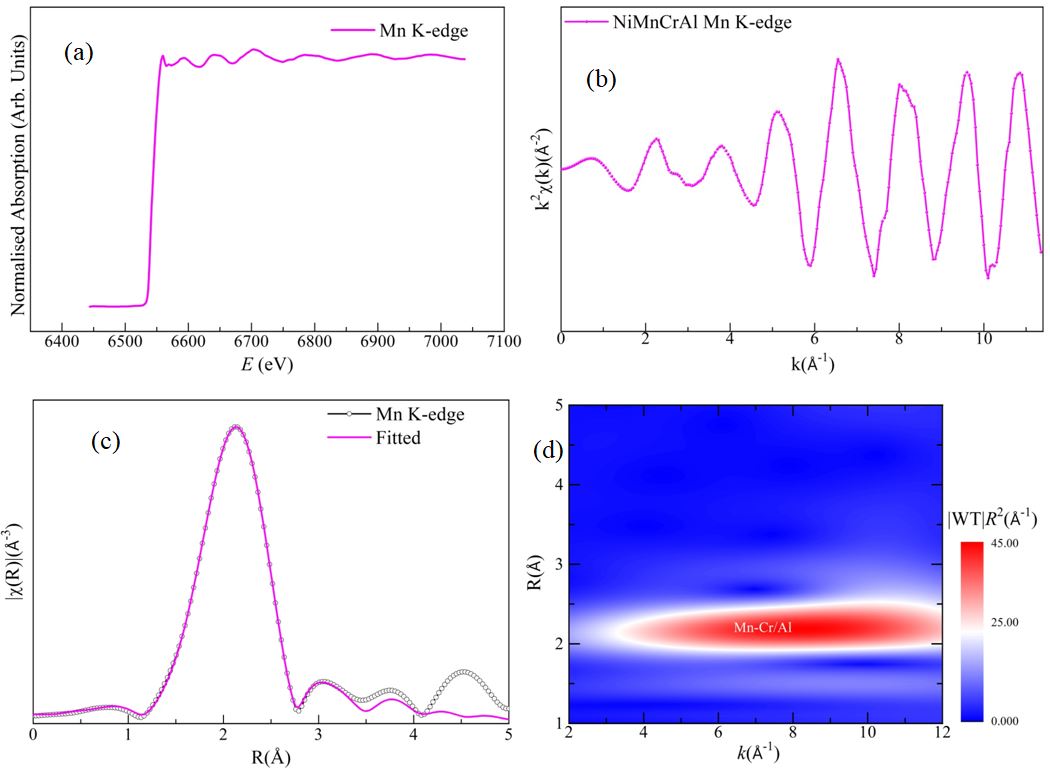}
	\caption{(a)--(d) show the corresponding EXAFS results for the Mn $K$-edge: (a) normalized absorption spectrum, (b) $k^{2}\chi(k)$, (c) Fourier-transformed $|\chi(R)|$ with fit, and (d) WT-EXAFS contour plot, where the dominant contribution arises from the Mn--Cr/Al scattering path.}
	\label{fig:EXAFS-Mn}
\end{figure*}

The preceding measurements clearly establish a spin-glass feature in NiMnCrAl, rather than the long-range magnetic order expected in the presence of $L2_1$ disorder. This suggests that structural disorder plays a significant role in governing the magnetic behavior of the system. To investigate the nature and extent of this disorder, EXAFS measurements were carried out in two modes \textit{viz.} transmission mode and fluorescence mode, respectively.  In transmission mode, two ionization chambers (each 30 cm long) recorded the incident flux ($I_0$) and transmitted flux ($I_t$). The absorption coefficient was obtained as $\mu(x) = \ln(I_0/I_t)$, where $x$ is the sample thickness along the beam direction. Gas pressures and mixtures were chosen to give 10-20\% absorption in the first chamber and 70-90\% in the second, improving the signal-to-noise ratio. Beam energy was calibrated by setting the Mn K-edge to 6539 eV before the Mn K-edge XAS measurements using standard Mn-foil in transmission mode. The same procedure was followed for the Cr K-edge. Spectra at both Cr and Mn K-edge for the sample were recorded in fluorescence mode using a silicon drift detector for the fluorescence intensity ($I_f$), with an ionization chamber measuring $I_0$. Sample was mounted at $\sim 45^\circ$ to the incident beam. The region of interest was chosen from the fluorescence line of each edge. Here, the absorption coefficient was evaluated as $\mu = I_f/I_0$.

Background removal and fitting were done using Demeter, with Strawberry Perl software (version 0.9.25) \cite{EXAFS-2}. The EXAFS oscillations in $\mu(E)$ were converted to the absorption function after removing the bare atom background \cite{Fe2RuGe-Sudip}.

\begin{equation}
	\chi(E)=\frac{\mu(E)-\mu_0(E)}{\Delta\mu_0(E_0)}
	\label{eqn:EXAFS1}
\end{equation}

\noindent
where $E_0$ represents the absorption edge energy, $\mu_0(E)$ is the bare atom background, and $\Delta\mu_0(E)$ is the step in $\mu(E)$ value at the absorption edge. Further, this absorption function is converted to the wave number-dependent absorption coefficient, $\chi(k)$, using the following relation

\begin{table*}[ht!]
	\caption{EXAFS fitting parameters obtained from Mn and Cr K-edge analysis.}
	\begin{ruledtabular}
		\begin{tabular}{lccclccc}
			\multicolumn{4}{c}{\textbf{Cr edge}} & \multicolumn{4}{c}{\textbf{Mn edge}} \\
			\hline
			Path & R (\AA) & N & $\sigma^{2}$ & Path & R (\AA) & N & $\sigma^{2}$ \\
			\hline
			Cr-Ni & $2.818 \pm 0.024$ & 5.4  & $0.009 \pm 0.002$ & Mn-Cr & $2.447  \pm 0.019$ & 4.1  & $0.008 \pm 0.002$ \\
			Cr-Mn & $2.447 \pm 0.019$ & 3.9  & $0.008 \pm 0.002$ & Mn-Al & $2.494  \pm 0.019$ & 5.4  & $0.005 \pm 0.004$ \\
			Cr-Al & $2.582 \pm 0.076$ & 3.3  & $0.028 \pm 0.019$ & Mn-Ni & $2.444  \pm 0.017$ & 3.1  & $0.002 \pm 0.002$ \\
			Cr-Cr & $3.999 \pm 0.062$ & 10.8 & $0.020 \pm 0.004$ & Mn-Mn & $4.091  \pm 0.085$ & 10.8 & $0.034 \pm 0.014$\\
		\end{tabular}
	\end{ruledtabular}
	\label{tab:exafs}
\end{table*}

\begin{equation}
	k=\sqrt{\frac{2m(E-E_0)}{\hbar^2}}
\end{equation}

\noindent
where $m$ is the electronic mass, weighted by $k^2$ to enhance the oscillations at high $k$ values. Finally, for particular edge energy, a Fourier transform of $k^2\chi(k)$ is performed that generates the $\chi(R)$ versus $R$ plots displayed using Fig.~\ref{fig:EXAFS-Cr} \& \ref{fig:EXAFS-Mn}, where the plot in $R$-space exemplifies the Fourier transformation of the EXAFS signal and its peaks denote scattering amplitude of the photoelectron wave from the source atom to the neighboring absorber. These results were further visualized via Morlet wavelet transform (MWT) (used in Fig.~\ref{fig:EXAFS-Cr} \& \ref{fig:EXAFS-Mn}) as employed in the EvAX 5.1 code\cite{EXAFS-3}. The MWT-EXAFS is advantageous because it can able to resolve the scatterers in radial space (R-space) as well as in the wave vector space (k-space), simultaneously \cite{EXAFS-4}. MWT was carried out in the $k$-space range of 2--12 $\mathrm{\AA^{-1}}$, and an $R$-space range of 0-6 $\mathrm{\AA}$.

Element-selective local structure around Mn and Cr sites in NiMnCrAl was investigated using EXAFS analysis. Fig.~\ref{fig:EXAFS-Cr} \& \ref{fig:EXAFS-Mn} shows the normalized EXAFS spectra, $k^2$-weighted $\chi(k)$ oscillations, Fourier-transformed $\chi(R)$ with fits, and wavelet-transform (WT) plots for the Mn and Cr K-edges.The Fourier-transformed spectra at both edges are dominated by a single well-resolved nearest-neighbor peak, which was modeled using contributions from sum of single scattering paths \cite{EXAFS-5} as summarized in Table~\ref{tab:exafs}. At the Mn K-edge, the fit yields Mn-Ni, Mn-Cr, Mn-Al, and Mn-Mn bond distances of $2.444 \pm 0.017$, $2.447 \pm 0.019$, $2.494 \pm 0.019$, and $4.091 \pm 0.085$~\AA, respectively, with coordination numbers of 3.1, 4.1, 5.4, and 10.8. Importantly, the number of nearest neighbours were kept fixed during the fitting process so that no statistical uncertainties were imposed on these parameters. The R-factor came out to be 0.018 for both the R-space fitting of Mn and Cr K-edges. The corresponding Debye-Waller factors ($\sigma^2$) increase in the order Mn-Ni $<$ Mn-Al $<$ Mn-Cr $<$ Mn-Mn, indicating progressively larger static/thermal disorder along this sequence. At the Cr K-edge, the fit gives Cr-Mn, Cr-Al, Cr-Ni, and Cr-Cr distances of $2.447 \pm 0.019$, $2.582 \pm 0.076$, $2.818 \pm 0.024$, and $3.999 \pm 0.062$~\AA, with coordination numbers of 3.9, 3.3, 5.4, and 10.8, respectively. The Cr--Al shell exhibits the largest disorder parameter among the Cr-edge paths ($\sigma^2 = 0.028 \pm 0.019$~\AA$^2$), pointing to substitutional disorder on the Al sublattice in the vicinity of Cr. The Mn-Cr bond length obtained independently from the two edges ($2.447$~\AA\ at both) is in excellent agreement, validating the consistency of the fits. The relatively high $\sigma^2$ values associated with the Cr-Al and Mn-Mn paths suggest positional disorder involving the Al and Mn sublattices. The MWT-EXFAS analysis also supports these interpretations as for both the wavelet transformed EXAFS of Mn and Cr K-edge are able to visualize the diffuse nature of the scatterers at the higher shells in comparison to the strong coherent nature of the scatterers in the first shell. Furthurmore, this is also consistent with the antisite disorder inferred from the magnetic characterization of this compound.

\subsection{Neutron Diffraction}
\label{sec:Neutron}

\begin{table}
	\centering
	\caption{ The structure obtained from the Rietveld refinement of the powdered XRD and Neutron Diffraction pattern.}
	\label{tab:Actual Sctructure}
	\begin{tabular}{ccccc}
		\hline
		\hline
		Element & \hspace{1cm}  $ 4a $  & \hspace{1cm}  $ 4b $  & \hspace{1cm} $ 4c $ & \hspace{1cm} $ 4d $ \\ 		
		\hline
		Al  & \hspace{1cm} 0.4 & \hspace{1cm} 0.4 & \hspace{1cm} 0.1 & \hspace{1cm} 0.1 \\
		Mn  & \hspace{1cm} 0.4 & \hspace{1cm} 0.4 & \hspace{1cm} 0.1 & \hspace{1cm} 0.1 \\
		Ni  & \hspace{1cm} 0.1 & \hspace{1cm} 0.1 & \hspace{1cm} 0.4 & \hspace{1cm} 0.4 \\
		Cr  & \hspace{1cm} 0.1 & \hspace{1cm} 0.1 & \hspace{1cm} 0.4 & \hspace{1cm} 0.4 \\
		\hline
		\hline
	\end{tabular}
\end{table}

\begin{figure}
	\centering
	\includegraphics[width=8cm]{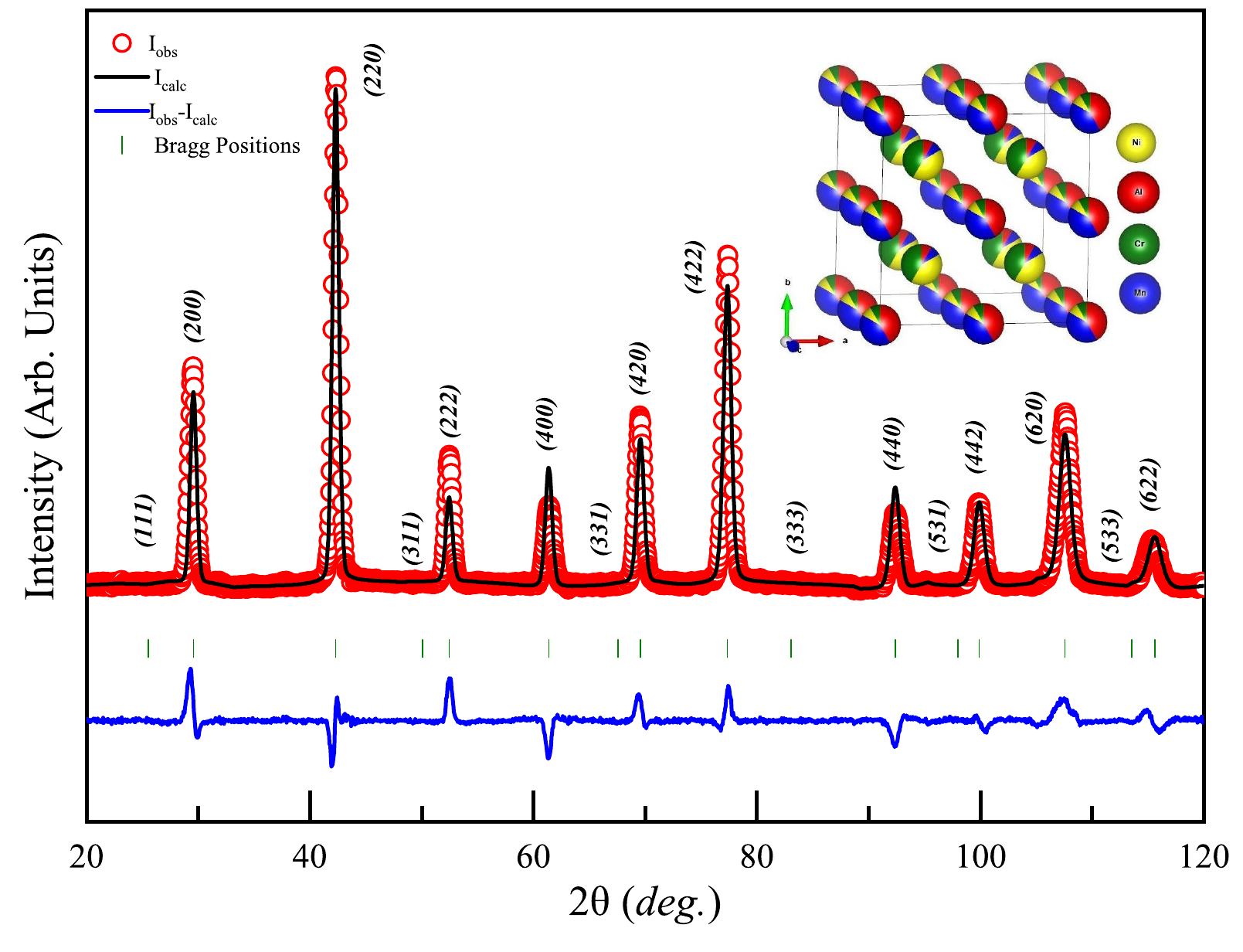}
	\caption{Rietveld refinement of the 300~K Neutron Diffraction (ND) data considering all the possible atomic intermixing as mentioned in the Table~\ref{tab:Actual Sctructure}. This structural model is consistent with the XRD analysis also. Inset: Structural model of the disordered NiMnCrAl}
	\label{fig:Neutron}
\end{figure}

As presented earlier in this work, the XRD pattern of NiMnCrAl could be well-explained by considering the $L2_1$ structure, whereas EXAFS analysis indicates the possible presence an additional degree of disorder, \textit{viz.}, $B2$-type atomic arrangement. Our theoretical calculation suggest that both the disorders would support long range ferrimagnetic ordering with magnetic moment of $\sim 1.6\,\mu_{\mathrm{B}}$/f.u. However, as our magnetic measurements unambiguously establish a cluster spin-glass state in low temperatures, one would tempt to associate the spin-disorder with atomic disorder far beyond $B2$-type. As Neutron scattering cross-sections are considerably more sensitive than x-ray scattering cross-sections for atoms with close atomic numbers, neutron diffraction (ND) measurement has been carried out at room temperature for investigating the atomic disorder in further detail.

Figure~\ref{fig:Neutron} shows the ND pattern recorded at 300~K. The complete absence of the $(111)$ superlattice reflection signifies site-mixing between crystallographically equivalent positions: specifically, exchange between both octahedral-octahedral and tetrahedral-tetrahedral sites. While the $L2_1$ disorder suggested by XRD pattern could detect only the swapping between octahedral-octahedral sites by Al and Mn atoms due to the relatively large difference between their atomic numbers, the additional tetrahedral-tetrahedral site disorder, representing $B2$-type, is indicated by EXAFS measurements. However, our attempt to explain the ND spectra using $B2$-type structure results in overestimation of (200) Bragg peak intensity which, in reality, found to be much smaller. This discrepancy indicates that the actual structure hosts a degree of disorder beyond both the $L2_1$ and $B2$ descriptions.

A satisfactory simulation of crystal structure was eventually obtained with a mixed-phase model comprising 80\% $B2$ and 20\% $A2$ disorder ($a = 5.79$~\AA), depicted in the inset of Fig.~\ref{fig:Neutron}. In this model, the site occupancies are distributed among all constituent atoms across all crystallographic positions. The octahedral sites are occupied predominantly by Al and Mn, with comparatively minor contributions from Cr and Ni, while the tetrahedral sites are occupied primarily by Ni and Cr, with smaller contributions from Al and Mn. The complete structural model, with estimated site occupancies, is presented in Table~\ref{tab:Actual Sctructure}. Notably, this same model was found to simultaneously describe the XRD pattern as well, lending further confidence to the proposed structural configuration.

\subsection{Theoretical Explanation}

\begin{figure}
	\includegraphics[width=8cm]{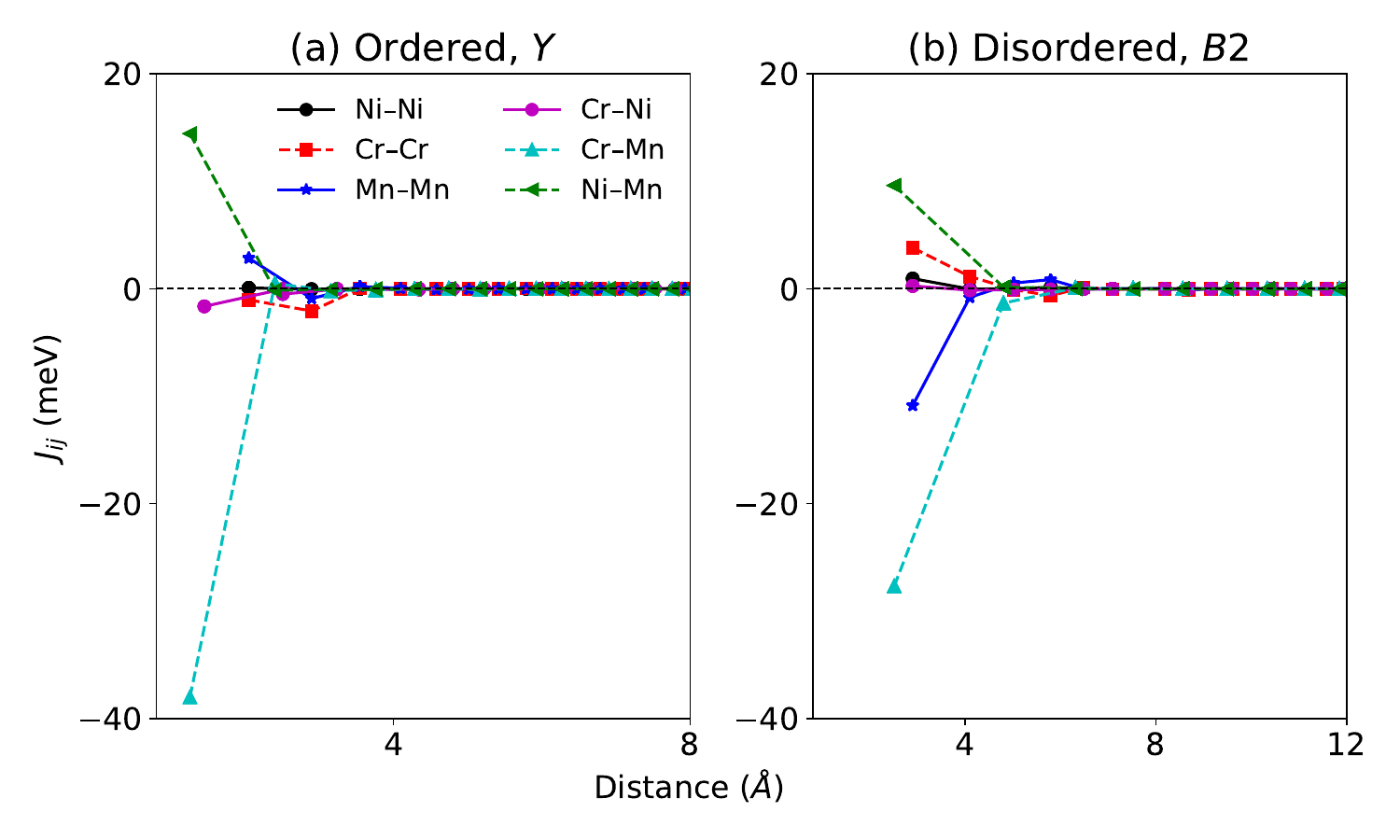}
	\caption{Heisenberg exchange coupling parameters of NiMnCrAl in (a) ordered, $Y$, and (b) disordered, $B2$ structures respectively.}
	\label{fig:Heisenberg interaction}
\end{figure}

To understand how the observed structural disorder influences the magnetic behaviour of NiMnCrAl, we calculated the Heisenberg exchange coupling parameters ($J_{ij}$) between the $i^{th}$ and $j^{th}$ magnetic atoms. In Fig.~\ref{fig:Heisenberg interaction} we present the $J_{ij}$ for the perfectly ordered $Y$ phase and the disordered $B2$ phase. In the ordered phase, the most dominant interaction is the antiferromagnetic coupling between Mn and Cr atoms, followed by a strong ferromagnetic Ni-Mn interaction, which together stabilize an overall ferrimagnetic ground state. In contrast, the intra-atomic interactions, such as Cr-Cr and Mn-Mn, are weak and exhibit a long-range oscillatory behaviour, owing to the large interatomic spacing (Mn-Mn or Cr-Cr) of $\sim$~4~\AA. This is consistent with the EXAFS results (Table~\ref{tab:exafs}), which give Cr-Cr and Mn-Mn distances of $3.999 \pm 0.062$~\AA\ and $4.091 \pm 0.085$~\AA, respectively.

In the case of the $B2$ structure, antisite disorder brings like atoms closer together, which results in a strong antiferromagnetic interaction between Mn atoms. The EXAFS results show that the second-neighbour Mn-Mn and Cr-Cr shells have large Debye-Waller factors ($\sigma^2 = 0.034 \pm 0.014$ and $0.020 \pm 0.004$, respectively), indicating a broad distribution of Mn-Mn and Cr-Cr distances and hence significant local disorder. It is evident from Fig.~\ref{fig:Heisenberg interaction} that when the Mn-Mn distance is reduced to $\sim$~2.9~\AA, a strong antiferromagnetic interaction between Mn atoms emerges, which is absent in the ordered structure. This additional direct Mn-Mn antiferromagnetic interaction, together with the Mn-Cr antiferromagnetic coupling, can introduce magnetic frustration and may ultimately lead to the formation of a spin-glass state.

It is to note that the above-mentioned $B2$ configuration, Ni and Cr, and Mn and Al were simultaneously swapped; \textit{viz.} the octahedral sites (tetrahedral sites) are equally shared by Mn and Al (Ni and Cr). However, there is small amount of intermixing between these sites as presented in Table-\ref{tab:Actual Sctructure} as well as Fig. \ref{fig:Neutron}(inset). This makes the magnetic interactions further complicated as the intra-atomic distances between Mn atoms for small percentages. It makes the magnetic ground states to be more susceptible to the local environment, and promotes spin-glass like magnetic ground state.

\section{Conclusions}

In summary, we have presented a combined theoretical and experimental investigation of the quaternary Heusler alloy NiMnCrAl, motivated by its predicted potential as a high-temperature spin-gapless semiconductor. Besides confirming the robust SGS character, our DFT calculations on the structurally ordered Y-type phase ($F\bar{4}3m$; no. 216) also predict an exceptionally high magnetc transition temperature ($\sim$1300~K), arising predominantly from Mn (octahedral, $4b$) and Cr (tetrahedral, $4c$) moments, with Ni contributing negligibly. Such a high $T_C$ would place NiMnCrAl among the very few non-Co-based Heusler alloys capable of sustaining the magnetic order with 100\% spin polarization well above 1000~K, underscoring its conceptual relevance to the broader understanding of exchange interactions in this class of materials.

Experimentally, polycrystalline NiMnCrAl was found to crystallize as a single-phase compound, although the Rietveld refinement of the powder XRD data could detect a $L2_1$-type ($Fm\bar{3}m$; no. 225) of atomic disorder between the $4a$/$4b$ Wyckoff sites. Our calculations for $L2_1$ sturcture show that such a disorder-induced enhancement of crystal symmetry would instead result in anhilation of SGS gap and consequent violation of Slater-Pauling rule, along with  reduced magnetic moment as well as T$_\mathrm C$. Our bulk magnetic measurements (ac-dc-susceptibility, isothermal magnetisation, ZFC-FC memory effect, and magnetic relaxation) however fail to detect even the theoretically proposed long-range magnetic order as the system freezes into a spin-glass state at $T_f \sim$~37~K. Neutron diffraction and EXAFS analysis resolved this discrepancy by revealing a more complex crystal structure where a dominant $B2$-type disorder is superimposed with moderate A2-type disorder. Our calculations show that this broad distribution of local magnetic environments across all the Wyckoff positions frustrates the exchange interactions sufficiently to suppress long-range order entirely, in favor of the glassy magnetic response.

Taken together, these results trace a complete pathway from a theoretically ideal, ordered SGS ferrimagnet with a very high ordering temperature, T$_\mathrm C$ $\sim$~1300~K for a non-Co based quaternary Heusler alloy to a real, chemically disordered crystal structure exhibiting spin-glass freezing below 37~K: a marked reduction of more than an order of magnitude in the relevant magnetic energy scale, largely attributed to atomic disorder. While disorder-induced suppression of T$_\mathrm C$ is a known phenomenon in Heusler alloys, the magnitude of this effect in NiMnCrAl and its role in erasing long-range order altogether is striking. This work highlights the critical importance of combining local (EXAFS) and long-range (ND/XRD) structural probes with theoretical investigation when evaluating disorder-sensitive functional properties predicted for Heusler-based spintronic candidates, and offers NiMnCrAl as a model system for studying the disorder-driven crossover between spin-gapless semiconducting and spin-glass ground states.

\begin{acknowledgments}
A.P. acknowledges the Saha Institute of Nuclear Physics (SINP) and the Department of Atomic Energy (DAE), Government of India, for financial support. P.D. wants to acknowledge DST/WISE-PDF/PM-78/2023(G) project. M.M. acknowledges the Indo-French Centre for the Promotion of Advanced Research (IFCPAR/CEFIPRA) for support under Project No.~6908-3.
\end{acknowledgments}

\end{document}